\documentclass[manuscript]{acmart}
\AtBeginDocument{%
  }

\usepackage{tabularx}
\usepackage{graphicx}    
\usepackage{subcaption}  
\usepackage{enumitem}
\usepackage{listings}
\usepackage{diagbox}
\usepackage[most]{tcolorbox}
\usepackage{array}
\usepackage{soul}
\usepackage[T1]{fontenc}
\usepackage[utf8]{inputenc}
\usepackage{newtxtext,newtxmath} 
\usepackage{textcomp}
\usepackage{tcolorbox}
\tcbuselibrary{listings, skins,breakable}

\newcolumntype{Y}{>{\raggedright\arraybackslash}X}
\definecolor{promptbg}{RGB}{248,248,248}
\definecolor{promptframe}{RGB}{210,210,210}
\usepackage{float}
\usepackage{xcolor}
\definecolor{ysgreen}{RGB}{46,125,50}      
\definecolor{ysorange}{RGB}{230,140,40}
\let\originalcite\cite%
\newcommand{\verifiedcite}[1]{\originalcite{#1}}

\renewcommand\footnotetextcopyrightpermission[1]{}
\setcopyright{none}
\usepackage{xcolor}

\newtcolorbox{promptbox}[1][]{
  enhanced,
  breakable,
  colback=gray!8,
  colframe=gray!50,
  boxrule=0.5pt,
  arc=2pt,
  left=8pt,right=8pt,top=6pt,bottom=6pt,
  fontupper=\small,
  before skip=6pt,
  after skip=6pt,
  #1
}
\lstdefinestyle{jsonstyle}{
  basicstyle=\small\rmfamily,
  breaklines=true,
  breakatwhitespace=false,
  columns=flexible,
  keepspaces=true,
  showstringspaces=false,
  tabsize=2,
  frame=none,
  aboveskip=0pt,
  belowskip=0pt,
}

\newtcblisting{jsonbox}{
  enhanced,
  breakable,
  listing only,
  listing style=jsonstyle,
  colback=gray!8,
  colframe=gray!50,
  boxrule=0.5pt,
  arc=2pt,
  left=8pt,right=8pt,top=6pt,bottom=6pt,
  before skip=6pt,
  after skip=6pt,
}

\author{Yusi Sun}
\authornote{These authors contributed equally to this work.}
\affiliation{%
  \institution{the University of Hong Kong}
  \city{Hong Kong}
  \country{China}
}
\email{soniasun@connect.hku.hk}

\author{Ying Jiang}
\authornotemark[1]
\affiliation{%
  \institution{University of California, Los Angeles}
  \city{Los Angeles}
  \country{United States}
}
\email{anajymua@gmail.com}

\author{Jiayin Lu}
\affiliation{%
  \institution{University of California, Los Angeles}
  \city{Los Angeles}
  \country{United States}
}
\email{jiayin.kay.lu@gmail.com}

\author{Jing Qian}
\affiliation{%
  \institution{Tongji University}
  \city{Shanghai}
  \country{China}
}
\email{jqian1590@tongji.edu.cn}

\author{Yin Yang}
\affiliation{%
  \institution{the University of Utah}
  \city{Salt Lake}
  \country{United States}
}
\email{yin.yang@utah.edu}

\author{Yong-Hong Kuo}
\authornote{Corresponding author.}
\affiliation{%
  \institution{the University of Hong Kong}
  \city{Hong Kong}
  \country{China}
}
\email{yhkuo@hku.hk}

\author{Chenfanfu Jiang}
\authornotemark[2]
\affiliation{%
  \institution{University of California, Los Angeles}
  \city{Los Angeles}
  \country{United States}
}
\email{cffjiang@math.ucla.edu}

\begin{document}

\title{Foresee: Making Procedural Futures Tangible through Generative Previews}

\renewcommand{\shortauthors}{Sun et al.}

\begin{abstract}
Procedural guidance can show users how to act without explaining what an action enables later. We present Foresee, an interface combining next-step guidance with consequence previews, method comparisons, and cross-step access. A hybrid pipeline combines structured rendering, image editing, and video generation. A literature-informed framework and six-participant formative study shaped its design. A 24-participant study compared AR guidance, AR with generated video, and Foresee across digital painting, gameplay, and cocktail making. Exploratory painting comparisons showed higher correct completion without a recorded error than AR and higher post-task quiz accuracy than generated-video guidance under task-wise correction. Interviews revealed that users could overlook valued explanations, match visible outcomes without understanding later possibilities, or wait for concrete demonstrations when symbolic cues were unclear. These findings inform generative previews that explain what actions enable or constrain while supporting ongoing execution.
\end{abstract}



\begin{teaserfigure}
  \includegraphics[width=\linewidth]{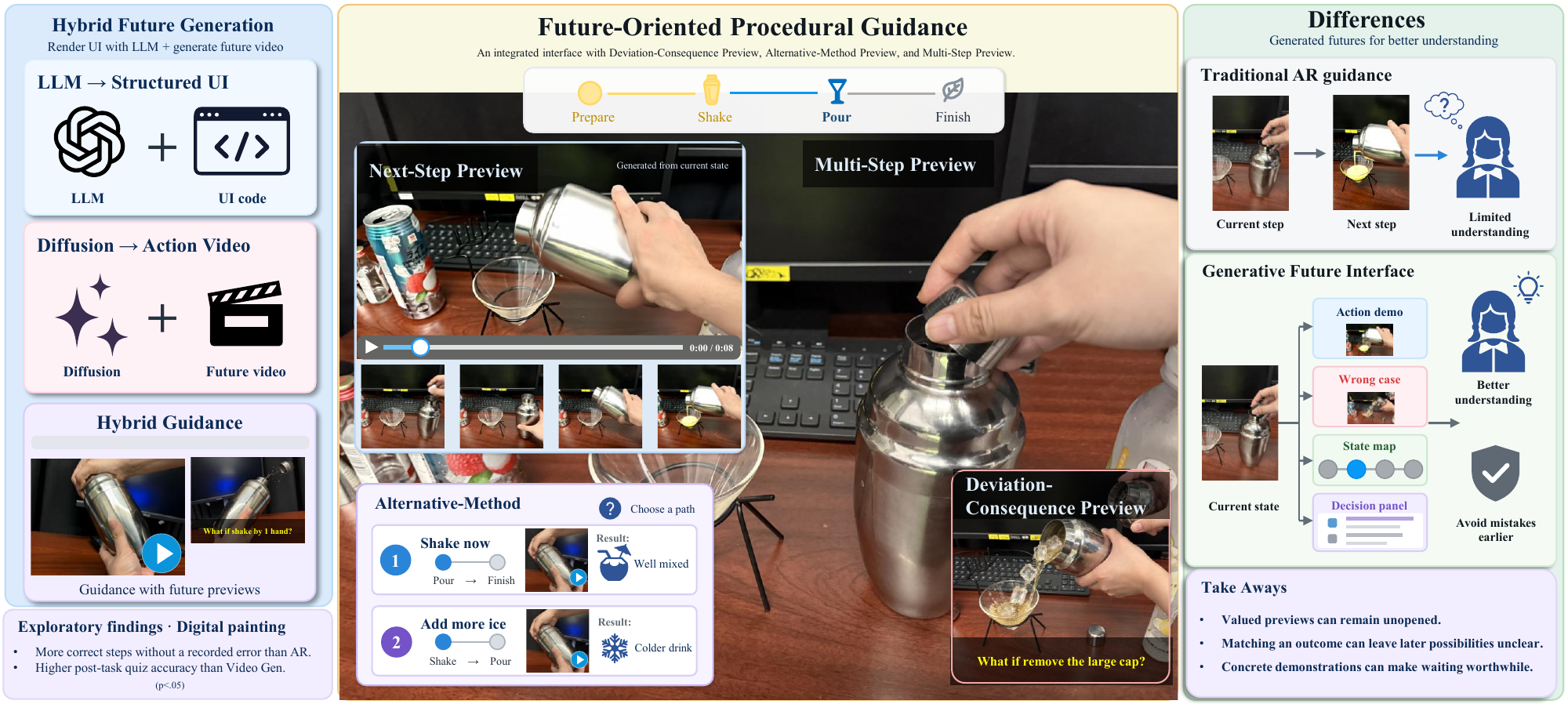}
  \caption{\textbf{Foresee: system overview and key findings.} The hybrid pipeline combines structured rendering, image editing, and video generation. Foresee combines ordinary next-step guidance with Deviation\textendash{}Consequence Preview, Alternative-Method Preview, and Multi-Step Preview. Exploratory Digital Painting comparisons and interview findings summarize task execution, post-task understanding, and how users engage with previews.}
  \Description{Three panels summarize Foresee. The left presents the hybrid generation pipeline. The center shows next-step guidance, deviation consequences, alternative methods, and access across steps. The lower-left panel summarizes exploratory painting comparisons. The right contrasts next-step guidance with future previews and summarizes interview themes: valued previews can remain unopened, matching an outcome can leave later possibilities unclear, and concrete demonstrations can make waiting worthwhile.}\label{fig:teaser}
\end{teaserfigure}

\maketitle

\section{Introduction}

Procedural guidance helps people carry out unfamiliar tasks through instructions and demonstrations~\verifiedcite{grossman2010toolclips}. Yet completing an instructed action does not necessarily explain why it matters or what it enables later~\verifiedcite{zhang2025reflective}. As an illustrative example, consider adding a sun to a digital landscape. A separate layer preserves independent movement; drawing on the background may look identical initially but later require erasing and redrawing other parts of the image. An instruction to create a layer leaves this value implicit. Connecting the action to its consequences could explain the dependency before users incur rework. Prior work offers complementary ways to make these relationships available. Feedforward interfaces expose possible effects before users commit to an action~\verifiedcite{muresan2023feedforward,coppers2019fortunettes}, and reflective AR prompts encourage users to consider action purposes and hypothetical alternatives~\verifiedcite{zhang2025reflective}. Multimedia learning provides a rationale for pairing explanatory words with relevant pictures~\verifiedcite{mayer2003promise}, while procedural tutorials such as MixT combine browsable text and images with per-step video to convey dynamic interactions~\verifiedcite{chi2012mixt}. Together, these approaches motivate guidance that explains an action, depicts its consequences, and supplies execution detail when needed.

Image and video generation offer a way to construct such representations from task specifications and available scene information~\cite{wan2025,kong2024hunyuanvideo}. We define \emph{generative future previews} as depictions of an upcoming action, possible subsequent state, or their relationship, given a starting state and candidate action or method. In the drawing example, an explanation and outcome image could show why moving a sun on a shared layer would disturb the background, while a video could demonstrate creating a separate layer. Generation produces task-relevant variants from shared procedural inputs and the current scene. We examine how users interpret and use these representations for prescribed actions and hypothetical alternatives. However, visibility alone does not make a future useful guidance. Users must interpret its connection to the current operation and decide whether further inspection is worthwhile. An image may clarify an expected appearance without explaining how the action changes later possibilities. A video may clarify motion or add redundant information and waiting. Alternative methods may enable meaningful comparison or leave users without a reason to prefer either route. Reflective AR prompts encourage reasoning about actions and outcomes, while Satori models user state and context to infer appropriate assistance~\verifiedcite{zhang2025reflective,satori25}. Foresee focuses on the representations through which prospective information becomes usable: pairing demonstrations of how to act with inspectable explanations of what those actions enable or constrain. The design challenge is to make these relationships interpretable and accessible within ongoing execution. This motivates our overarching question: \textbf{How can generative future previews make action--consequence relationships understandable within ongoing procedural guidance?} We address two research questions:
\begin{enumerate}
    \item \textbf{RQ1:} How do users interpret and use generative future previews during procedural tasks, and what information and forms of access do they find useful?
    \item \textbf{RQ2:} How do AR guidance, AR with generated video, and Foresee differ in execution outcomes and post-task procedural understanding?
\end{enumerate}

To address RQ1, we developed a literature-informed framework and explored six preview configurations in an iterative formative study with six participants. Feedback favored directly visible consequences over separate exploration controls, shaping Foresee's combination of ordinary next-step guidance, consequence previews, method comparisons, and cross-step access (Figure~\ref{fig:teaser}). A hybrid pipeline instantiates these previews from reviewed procedural inputs and current scene images.

To address RQ2, a 24-participant study compared the three guidance conditions within digital painting, gameplay, and cocktail making, measuring execution and post-task understanding. Interviews extended RQ1 through accounts of preview use and interpretation.
Exploratory task-specific analyses identified benefits in Digital Painting. Foresee supported higher correct completion without a recorded error than AR alone ($p<.05$) and higher quiz accuracy than Video Gen ($p<.05$). Participant accounts suggest that concrete demonstrations can help users interpret abstract action cues, while consequence previews need clear connections to the current task.

This work makes three contributions:
\begin{enumerate}
    \item \textbf{A framework and interface for inspectable procedural futures.} A literature-informed framework and formative refinement (N=6) shape Foresee's combination of consequence previews, method comparison, and cross-step access.
    \item \textbf{A hybrid generation pipeline for physical and GUI-based guidance.} Structured action--outcome specifications guide the generation of videos depicting prescribed actions and their intended outcomes. Combining UI rendering, image editing, and video generation supports both physical operations and GUI interactions.
    \item \textbf{An evaluation of Foresee's support for task execution and procedural understanding.} A three-condition study ($N=24$) across three tasks compares Foresee with AR and generated-video guidance, assessing execution and post-task understanding. Participant accounts explain preview use and inform design recommendations.
\end{enumerate}

\section{Related Work}
\subsection{Situated Guidance and Prospective Representations}

Procedural guidance systems seek to connect instructions to their execution environment. Situated AR systems place text, labels, arrows, target poses, and animations on or near relevant objects. Studies in maintenance and assembly show that such in-situ presentation can improve localization, accuracy, completion time, or workload relative to detached or side-by-side documentation, although the benefit depends on registration quality and visualization design~\cite{henderson2011documentation,henderson2011psychomotor,blattgerste2018insitu,jasche2021visualizations}. Authoring and retargeting systems broaden this representational vocabulary: ProcessAR and TutorialLens capture demonstrations as stepwise 2D/3D guidance, PaperToPlace spatializes instruction documents, and media-retargeting pipelines recover registered object motion, target configurations, or tool trajectories from documentation and video~\cite{processar21,tutoriallens21,papertoplace23,retarget15,videoar17,videoar20,ihara2025video2mr}. These systems make procedures actionable in context but primarily guide users along authored routes.

Guidance can also respond to task state and user progress. Context-aware AR systems track assembly status and update instructions during execution~\cite{khuong2014contextaware,adaptuar21}, while demonstration-based systems link guidance to spatial, temporal, or progress triggers~\cite{tutoriallens21,processar21,instruar23}. More recent AI assistants use scene observations and task progress to contextualize assistance: XaiR grounds responses in egocentric observations and spatial mapping, Guided Reality selects and places visual guidance, and Vid2Coach provides progress-aware instructions and feedback~\cite{xair24,guided25,video2coach25}. Satori uses user intent and environmental context to guide interventions, while AdaptiveCoPilot adjusts modality and information load using behavioral and workload signals~\cite{satori25,wen2025adaptivecopilot}. These approaches inform how guidance can be integrated into ongoing activity.

Feedforward research more directly treats a not-yet-executed interaction as an interface object. In graphical interfaces, Fortunettes computes and visualizes a widget's possible future availability or value before commitment~\cite{coppers2019fortunettes}. In immersive environments, feedforward has been characterized through the triggering, previewing, and exiting of prospective action and outcome representations~\cite{muresan2023feedforward}. XR systems instantiate these representations with dynamic gesture paths, trajectories, target poses, ghosted hands or bodies, animation, and contextual video~\cite{fennedy2021octopocusvr,feedforward24,tutorivr19}. The principal unit of preview in this line of work is an interaction possibility and its immediate execution or outcome.

Several systems move beyond a single immediate preview. ViRgilites provides multilevel feedforward for alternative interaction possibilities, and BeatriXR supports the design and comparison of feedforward configurations~\cite{artizzu2024virgilites,artizzu2026beatrixr}. MR-Driven Near-Future Realities explores the preview and reflection of possible everyday scenarios~\cite{mathis2024nearfuture}. Building on these representations, we examine how consequence previews and action demonstrations fit into ongoing procedures.

\subsection{Supporting Understanding During Procedural Guidance}

Procedural understanding concerns how an action changes the task state and why that change matters for subsequent steps or goals. Research distinguishes performance during instruction from learning and transfer, motivating attention to what users understand as well as what they complete~\cite{Nicholas_15_Learning_Performance,Eiriks_11_Procedural_Instructions}. Kieras and Bovair showed that knowledge of component and control relationships helped users infer and simplify operating procedures~\cite{KIERAS1984255}. In the present work, consequence understanding concerns these action--state--goal relationships, including adverse consequences of deviations and differences between feasible methods. Feasible methods need not involve mistakes: they may share an immediate goal but differ in later implications. We assess post-task procedural understanding alongside execution.

Interfaces can support these relationships through explanation, reflection, and comparison. Self-explanation research demonstrates benefits from prompting learners to integrate new information with existing knowledge~\cite{CHI1994439}. In AR-guided tasks, \emph{From Following to Understanding} uses reflective prompts to connect actions with outcomes and consider hypothetical scenarios, reporting higher understanding quiz scores and more proactive information seeking~\cite{zhang2025reflective}. Example-based learning provides a complementary rationale for presenting cases that users can interpret: Schwartz and Bransford found benefits for subsequent prediction from analyzing contrasting cases before an explanation~\cite{Schwartz01121998}, while Adams et al.\ reported delayed learning benefits from identifying, explaining, and correcting erroneous solutions in mathematics tutoring~\cite{ADAMS2014401}. These findings motivate examining how brief previews support inspection and comparison of action--consequence--goal relationships during execution.

Reflective AR establishes the value of prompting action--outcome reasoning during execution~\cite{zhang2025reflective}. Foresee investigates how generated depictions and accompanying explanations make those relationships inspectable: a recognizable result can support an immediate check while leaving its implications for later actions unclear.

\subsection{Generative Visual Guidance and Controllability}

AI-based systems increasingly automate the construction of guidance artifacts. TutoAI extracts components for mixed-media tutorials, TAGGAR localizes reusable action visualizations, and CARING-AI generates context-aware avatar instructions~\cite{chen2024tutoai,TARGAR24,caring-ai25}. ARify transforms narrated videos into editable AR tutorials, RePairAR derives reassembly guidance from disassembly recordings, and agentAR generates executable AR applications from natural-language requests~\cite{hu2026arify,deng2026repairar,agentar25}. Beyond extracting, grounding, and constructing instructions, depicting actions and outcomes requires control over generated content.

Action-conditioned generation depicts how a current scene might change through an interaction. CosHand synthesizes a post-interaction image from a scene and a desired hand interaction, while InterDyn generates a temporally extended interaction from an initial frame and a motion control signal~\cite{coshand24,interdyn25}. Applying these capabilities to procedural guidance requires consistent object identity, target location, action order, and resulting state. Evaluations such as T2VPhysBench, Physion-Eval, and MechVerse identify challenges in generating physically and mechanically consistent motion~\cite{t2vphysbench25,physioneval26,mechverse26}.

HCI systems improve generative controllability by externalizing constraints. Map2Video grounds actor placement, trajectories, and camera framing in map and street-view geometry, while VideoCraft uses mixed-reality spatial layers to localize style and geometry transformations~\cite{jo2025map2video,li2025videocraft}. These creative-authoring techniques align generation with spatial intent. Foresee connects such control to instructional requirements: actions must also be correct, feasible, and appropriate to enact.

Explicit structure offers a complementary route to reliability. ElastoGen incorporates elastodynamic equations into learned 4D generation, while MotionCraft and Physical Simulator In-the-Loop Video Generation use simulated motion to guide generative video synthesis~\cite{feng2026elastogen,motioncraft24,psivg26}. In interfaces, formal application or task structure can avoid unconstrained pixel prediction altogether: Fortunettes computes immediate GUI futures from event handlers, Demo2Tutorial reconstructs hierarchical task graphs from demonstrations, and agentAR produces executable AR application structures~\cite{coppers2019fortunettes,bai2026demo2tutorial,agentar25}. Structure makes transitions inspectable and controllable where objects, actions, and state changes can be represented explicitly. Pixel generation covers broader appearances and physical actions but permits more plausible-looking, procedurally invalid transitions.

These approaches motivate combining explicit task constraints with generated imagery and motion for procedural guidance.

\section{Formative Study}

To address \textbf{RQ1}, we combined a literature-informed framework, six candidate configurations, and a pilot study to explore preview content, access, and representation during task execution. Section~\ref{sec:final-interface-scope} summarizes the resulting design.

\subsection{A Literature-Informed Design Framework}
We conducted a purposive literature search on procedural guidance, feedforward, adaptive assistance, and prospective visualization. An initial keyword search in the ACM Digital Library, Google Scholar, SpringerLink, and IEEE Xplore was narrowed to 11 core papers selected for their relevance to our design question. Starting from these papers, we combined further targeted searches with citation chasing~\verifiedcite{wohlin2014snowballing}, adding 34 papers in the second round and 26 in the third, for a corpus of 71 unique papers. Repeated retrieval of previously encountered papers motivated the pragmatic decision to stop after three rounds. The corpus informed the design vocabulary used to develop our prototypes.

We included task-executor-facing representations of upcoming actions, target states, or conditional outcomes connected to a physical or digital task. These comprised runtime guidance and separately described learner-facing outputs of authoring or generation systems. We distinguished these outputs from author-only controls, current-state feedback, and backend prediction or generation. Adjacent work informed the design vocabulary, while learner-facing configurations supplied the interface examples. Recorded demonstrations were eligible when their content concerned the learner's upcoming action.

One researcher coded interface implementations, representations, and interactions. Each round, they discussed a provisional dimension table with two group members uninvolved in prototype design or literature coding and synthesis, then used their feedback to clarify definitions and overlapping categories. Configurations specify user role, research stage, and interface condition, situating components within their source systems. Four dimensions organize this vocabulary: \textit{Future Semantics, Future Structure, Future Interaction, and Future Representation}. Their fields describe, compare, and combine design properties.

We distinguish source-described configurations, our analytical interpretations, and prototype design choices. Unspecified source properties are recorded as unknown.

\textbf{Future Semantics (S)} describes the assumed state and action, the depicted content, and its relation to a task goal. It distinguishes, for example, showing a target location from conveying why a deviation matters.

\textbf{Future Structure (T)} describes the inspectable unit, organization, horizon, and relations between futures. A sequence of planned steps differs from a comparison of methods toward a shared goal.

\textbf{Future Interaction (I)} describes how content is proposed, accessed, edited, and navigated, and whether a choice advances the guidance. It separates optional inspection from required branch selection and preview navigation from task progression.

\textbf{Future Representation (R)} describes the medium and its spatial, visual, and temporal presentation. AR anchoring is independent of medium: an upcoming action can be conveyed through a spatial cue, an image, or a video. A consequence image makes a possible state inspectable, but understanding why it matters also depends on the task context communicated alongside it.

These dimensions separate content from inspection: extending the horizon changes structure, whereas changing viewpoint changes representation.

\subsection{Selecting Candidate Configurations for Prototype Exploration}\label{sec:formative-candidates}

We selected six configurations through literature-informed developer judgment, considering relevance to procedural guidance and feasibility within the shared pipeline. The framework articulated their differences and informed this selection.

The selected set addresses three distinctions relevant to RQ1: continuing along a prescribed procedure, inspecting a hypothetical deviation, and recovering from an actual departure; comparing proposed methods and adjusting the current action through predefined options; and inspecting the immediate step and a broader procedural future. We explored these complementary functions within one prototype, bounded by available state information, reviewed task specifications, supported actions, generation waits, and content review.

\begin{figure}[htbp]
  \centering
  \includegraphics[width=\linewidth]{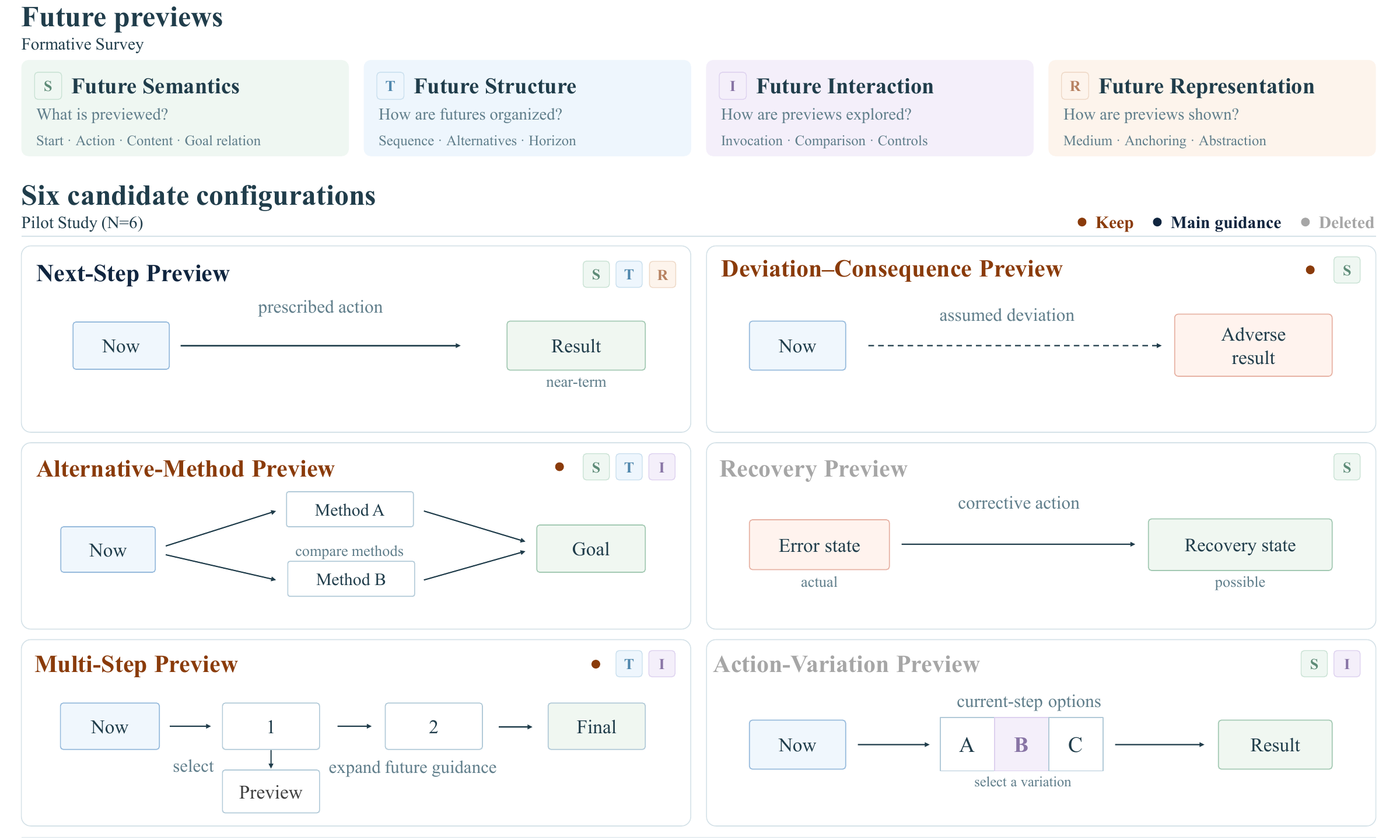}
  \caption{\textbf{Four design dimensions and six initially explored configurations.} Letter badges identify each candidate's emphasized dimensions. Title colors indicate the final roles: reddish-brown for retained preview components, dark blue for main next-step guidance, and gray for candidates excluded from the final interface (Section~\ref{sec:final-interface-scope}).}
  \Description{Four dimension headings--Semantics, Structure, Interaction, and Representation--sit above six schematics. Next Step links the prescribed action to a near-term result; Deviation\textendash{}Consequence links an assumed deviation to an adverse result; Alternative Method compares paths toward a goal; Recovery links an actual error state to corrective action; Multi-Step shows selecting a step to expand its preview; and Action-Variation offers predefined current-step options. The Next-Step title is dark blue, indicating main guidance. Deviation-Consequence, Alternative-Method, and Multi-Step titles are reddish-brown, indicating retained preview components. Recovery and Action-Variation titles are gray, indicating exclusion from the final interface.}\label{fig:future-previews}
\end{figure}

The initial candidate set included: \emph{Next-Step Preview}, \emph{Deviation\textendash{}Consequence Preview}, \emph{Alternative-Method Preview}, \emph{Recovery Preview}, \emph{Multi-Step Preview}, and \emph{Action-Variation Preview}. Figure~\ref{fig:future-previews} summarizes their relationships using the next prescribed action and its near-term result as a common descriptive reference. In the exploratory prototype, each configuration had an appropriate interaction: alternatives were compared, Multi-Step Preview expanded to show future guidance, and action variations were selected through predefined current-step controls. Section~\ref{sec:final-interface-scope} summarizes the retained candidates and their final forms.

Next-Step Preview, Deviation\textendash{}Consequence Preview, Alternative-Method Preview, and Recovery Preview primarily distinguish action conditions and goal relations; Multi-Step Preview changes organization and navigation, and Action-Variation Preview changes how users specify the current action. Their cross-dimensional organization supports combination: a map could organize alternative methods or wrong-case consequences, while an edited action could generate a valid variation or an undesirable result. Interaction and representation determine how users access and inspect each configuration.

\paragraph{Distinguishing prospective consequences from corrective feedback.}
Deviation\textendash{}Consequence Preview connects a relevant state, a specified hypothetical deviation, a possible consequence, and the task requirement that makes the consequence matter. For example, drawing a sun on the background layer becomes problematic when the task requires later moving it independently. A brief description of that condition and an outcome image can make the potential loss of editability or repair cost visible before the user acts. Alternative-Method Preview instead compares feasible routes to a shared goal. The initial prototype also included a candidate labeled \emph{Recovery Preview}, accessed through \emph{What Can I Do Now}, which supplied corrective guidance after an actual departure from the expected state. We initially explored this corrective function alongside prospective guidance.

\paragraph{Task-specific, constrained action editing.}
Action-Variation Preview generated consequence videos for predefined adjustments to the current action: brush color, size, and opacity in digital painting; the object to collect in gameplay; and ice or ingredient quantities in cocktail preparation. These numeric parameters and categorical targets varied the action while preserving the goal and starting state.

\paragraph{Instantiating configurations in a shared pipeline.}
The six candidates shared a workflow: establish inputs and task requirements, specify action conditions, generate or organize content, and present it. This pipeline accommodated parameter changes, ordering changes, and downstream effects. The task requirements determine whether a parameter change expresses a Wrong Case, a valid alternative, or a trade-off.

Following \emph{From Following to Understanding}~\verifiedcite{zhang2025reflective}, we phrased brief cues as questions to encourage curiosity and consideration of consequences.

Generation cost shaped presentation. We kept required Next Step guidance focused on the current action, with ordinary action videos providing execution detail. Outcome images conveyed states or consequences, and further action videos supplied process information when needed. Invocation, generation, and playback were configured separately: a user request could trigger a clip that then played automatically. This organization aimed to limit additional information and waiting in the mandatory guidance path.

\subsection{Pilot Study}\label{sec:formative-pilot}

We explored the six candidates with six participants using think-aloud tasks, questionnaires, and interviews. A prerecorded-video comparison condition helped participants reflect on the usefulness of generated guidance and additional previews. Iteration brought valued but rarely opened information into view and set aside burdensome controls; feedback informed the final selection.

\begin{figure}[H]
  \centering
  \begin{subfigure}[b]{0.32\linewidth}
    \centering
    \includegraphics[width=\linewidth]{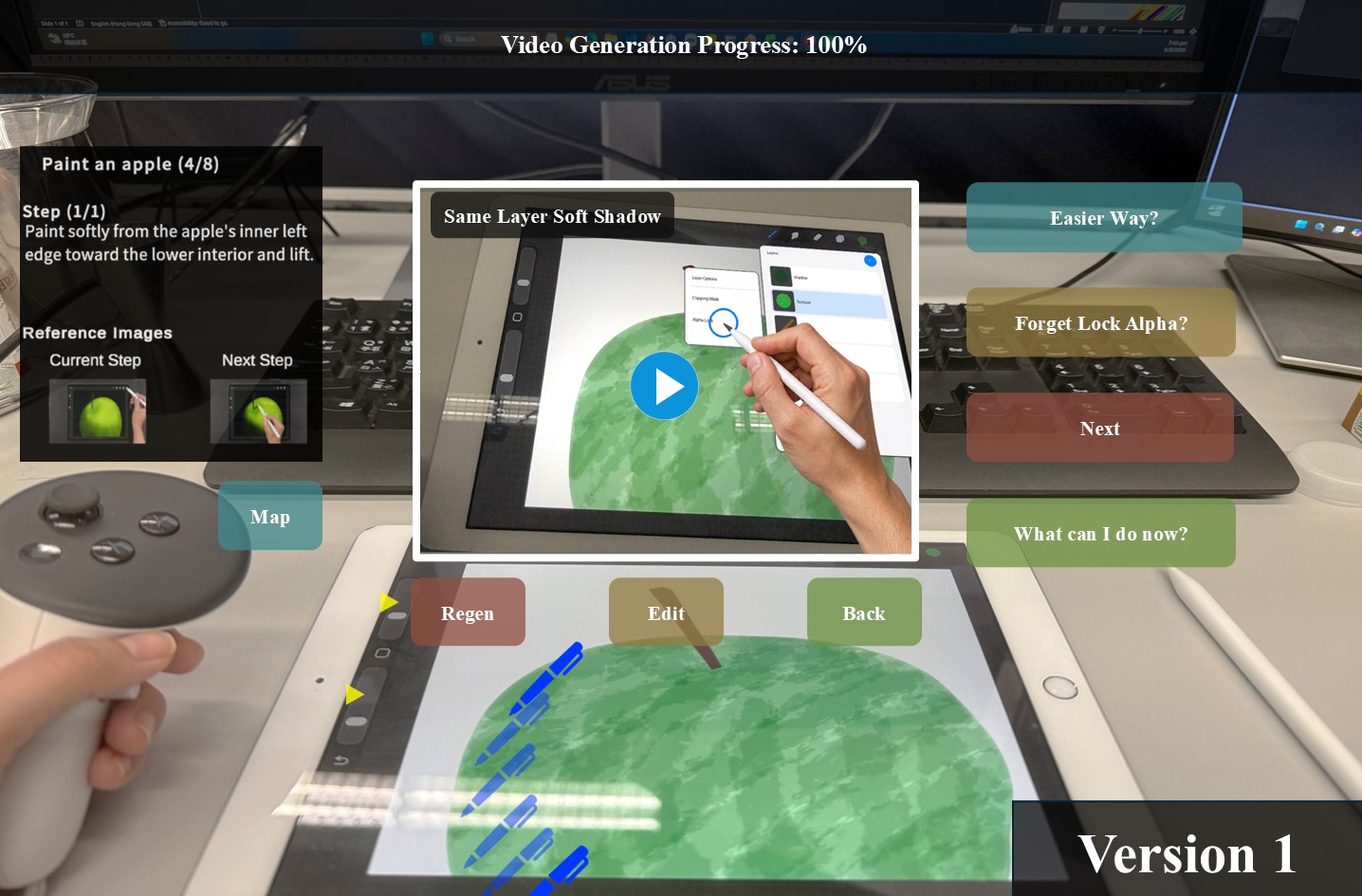}
    \caption{V1}\label{fig:pilot-ui-v1}
  \end{subfigure}\hfill
  \begin{subfigure}[b]{0.32\linewidth}
    \centering
    \includegraphics[width=\linewidth]{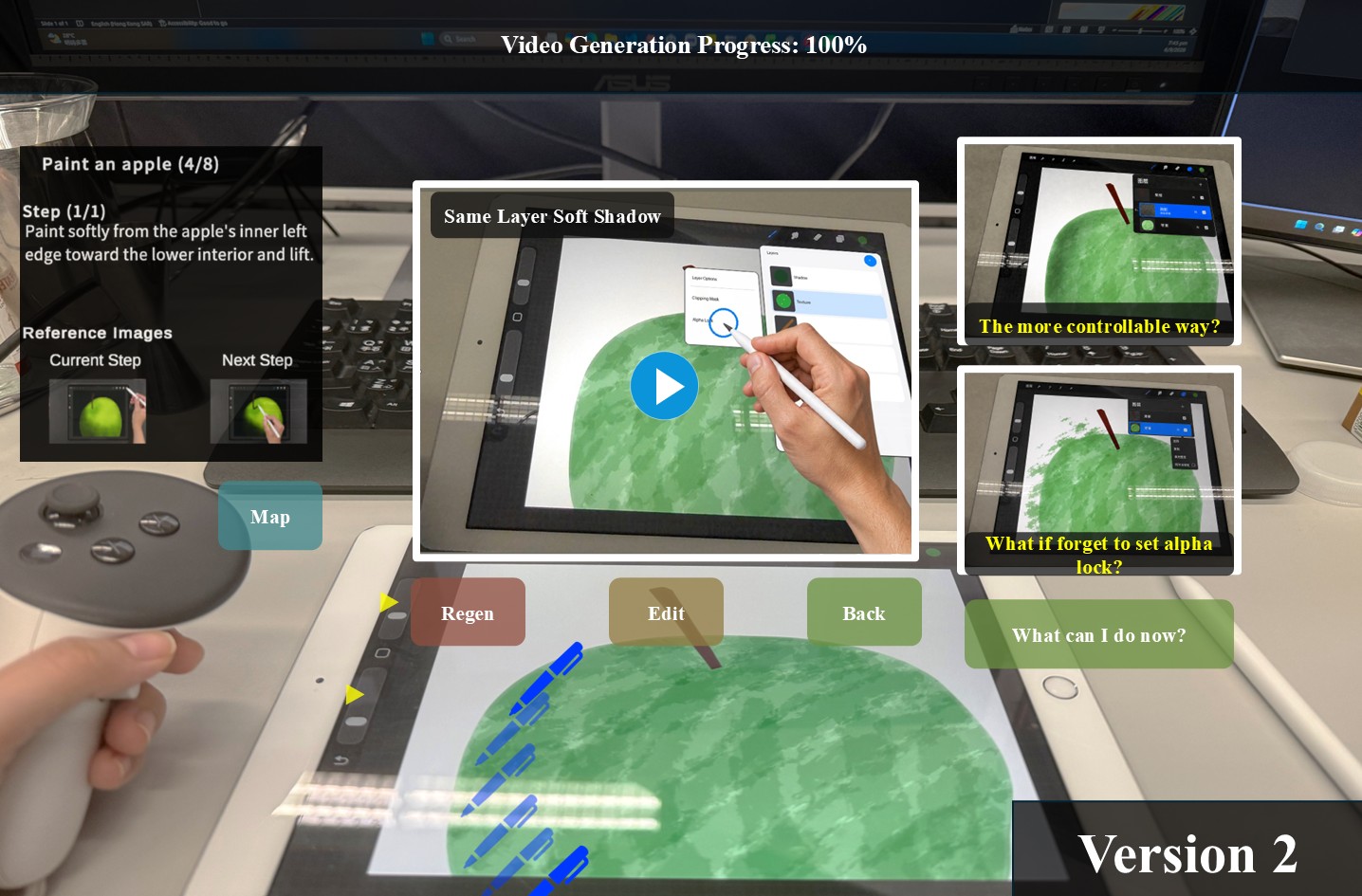}
    \caption{V2}\label{fig:pilot-ui-v2}
  \end{subfigure}\hfill
  \begin{subfigure}[b]{0.32\linewidth}
    \centering
    \includegraphics[width=\linewidth]{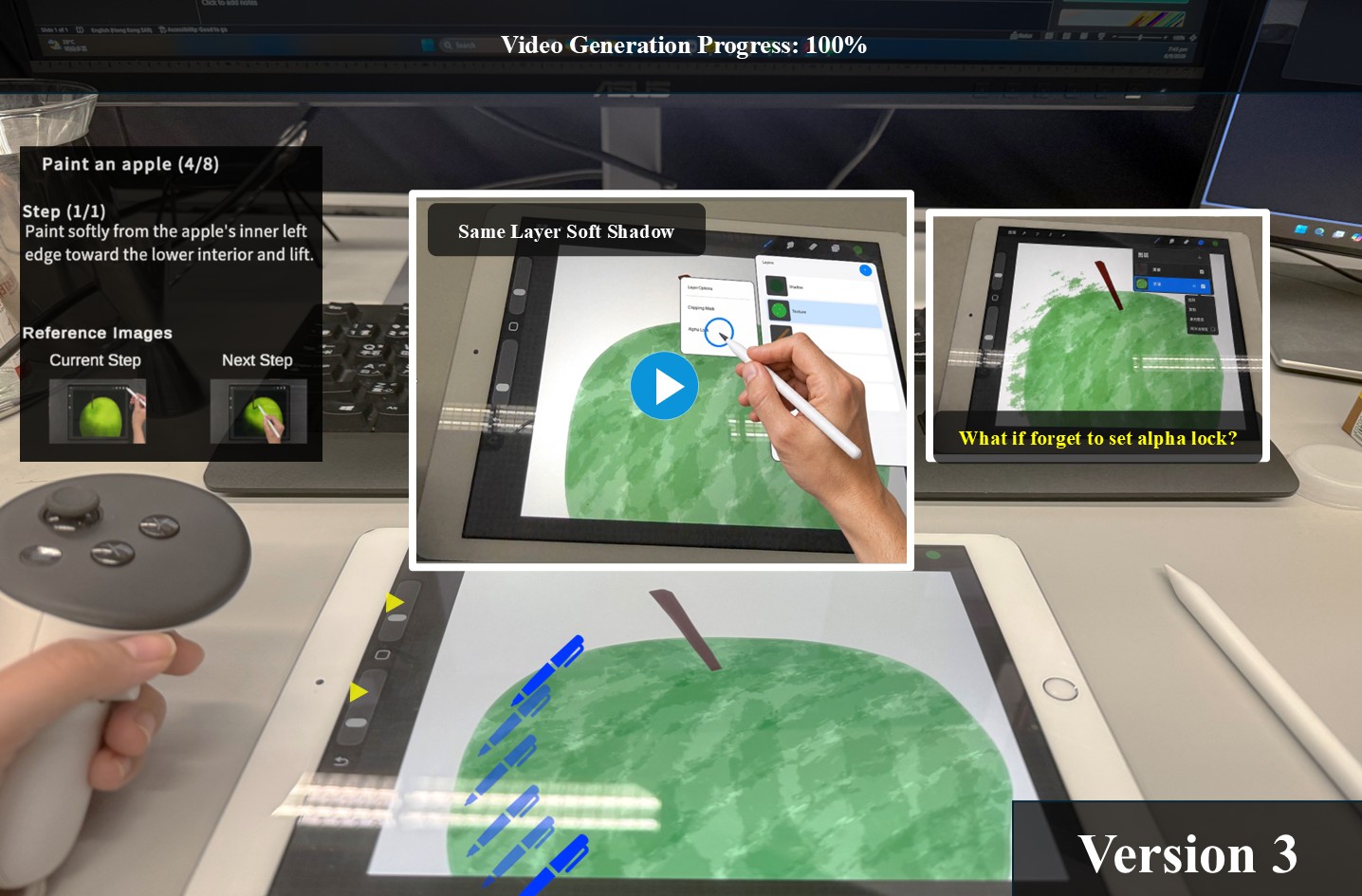}
    \caption{V3}\label{fig:pilot-ui-v3}
  \end{subfigure}
  \caption{Interface evolution during the formative study. V1 provides multiple controls for accessing additional guidance. V2 directly displays alternative and wrong-case previews. V3 removes additional controls, exposes Multi-Step Preview, and retains a directly visible wrong-case preview alongside the action-video panel. These screenshots document formative revisions; the final selection is summarized in Section~\ref{sec:final-interface-scope}.}\label{fig:pilot-ui-evolution}
  \Description{Three side-by-side screenshots show successive AR interface versions over the same tablet-painting task. V1 has an instruction panel, a central video panel, and multiple buttons. V2 replaces the alternative and wrong-case buttons with image previews. V3 displays Multi-Step Preview at the top, keeps the instruction and video panels and a wrong-case image, and removes the surrounding extra buttons.}
\end{figure}

We revised the prototype after each participant. V1, V2, and V3 denote major milestones used by P1--P2, P3--P4, and P5--P6, respectively, with smaller changes within each pair (Figure~\ref{fig:pilot-ui-evolution}). Later revisions introduced Qwen Image generation for outcome previews.

\subsubsection{Participants.}
We recruited six participants from a local university (P1--P6; four male, two female; age: $M=22$, $SD=4.04$ years). P1 reported familiarity with gameplay and P5 with cocktail preparation; the others reported no prior experience with the study tasks. One participant had previously used a Quest 3.

The study received university ethics approval, with identifying details omitted for anonymous review. Participants did not consume the alcoholic drinks they prepared, and headset discomfort prompted a pause and rest.

\subsubsection{AR Guidance System.}
The Quest 3 prototype provided step instructions, reference images, AR action cues, and video. The comparison condition used prerecorded videos; the generative condition used runtime-generated videos and the additional previews available in each revision. Both shared playback controls and automatic state-triggered progression in gameplay, with manual advancement in the other tasks.

From V3 onward, the generative condition included two fixed method decisions per task requiring a choice before guidance advanced. Other preview inspection was optional. Both conditions used backend GUI-target assistance when needed in gameplay and painting, without spoken or pointing assistance. Cocktail preparation used no such assistance.

Generated previews used the current scene image and a predefined step prompt that a language model could adapt. Failed attempts were retried.

\subsubsection{Tasks.}
\begin{figure}[htbp]
  \centering
  \begin{subfigure}[b]{0.32\linewidth}
    \centering
    \includegraphics[width=\linewidth,height=0.75\linewidth]{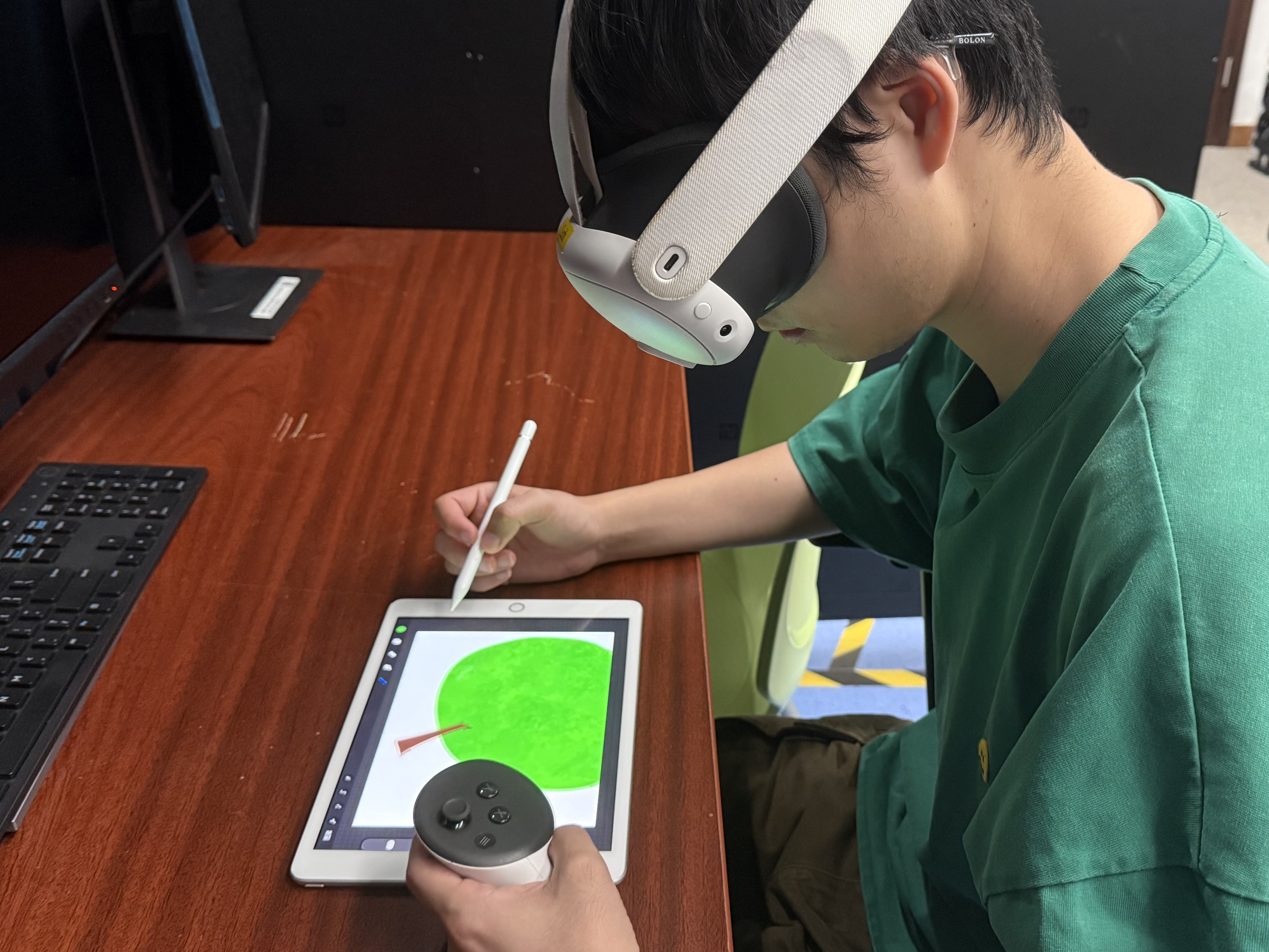}
    \caption{Painting}
  \end{subfigure}\hfill
  \begin{subfigure}[b]{0.32\linewidth}
    \centering
    \includegraphics[width=\linewidth,height=0.75\linewidth]{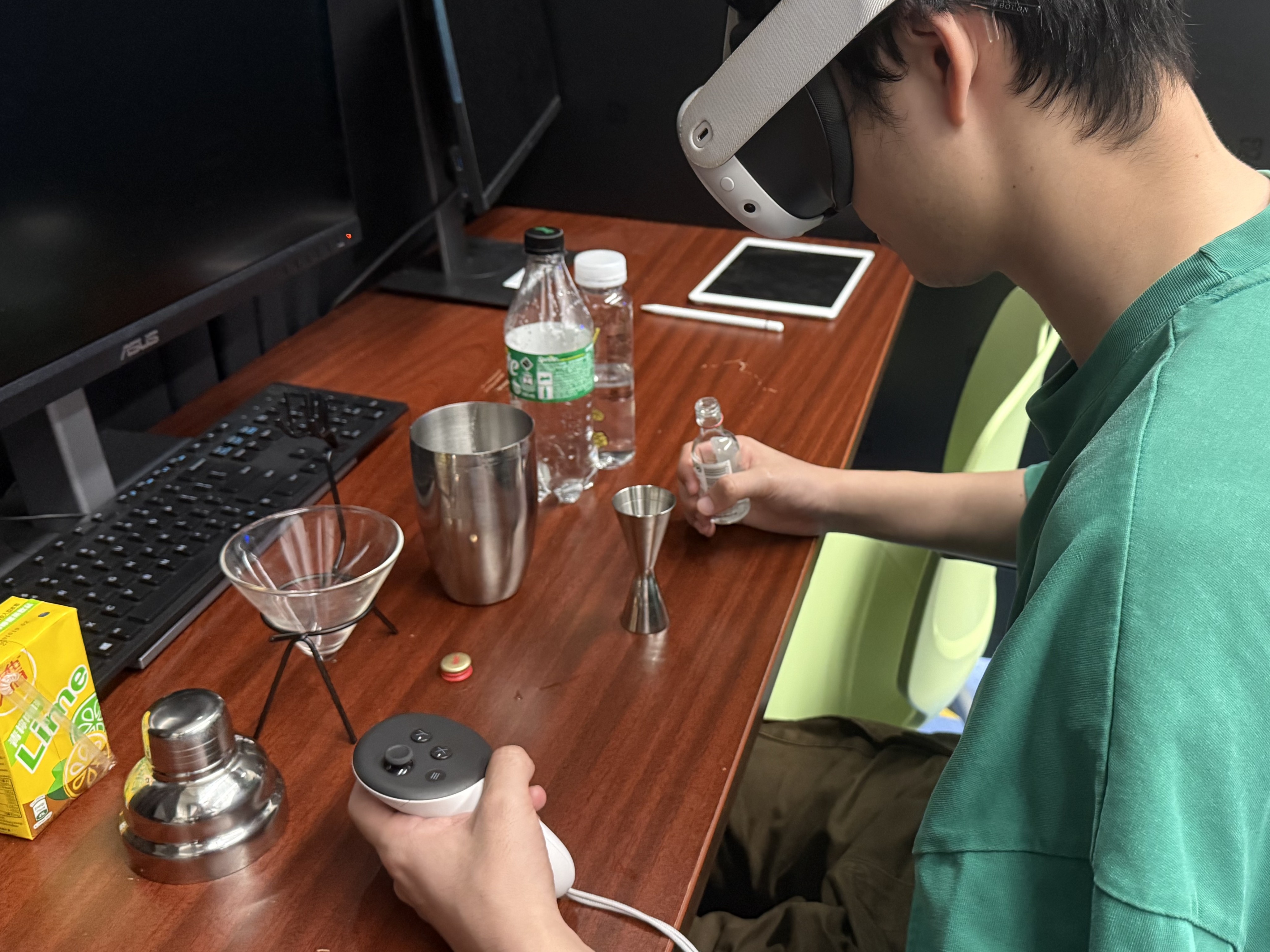}
    \caption{Cocktail Making}
  \end{subfigure}\hfill
  \begin{subfigure}[b]{0.32\linewidth}
    \centering
    \includegraphics[width=\linewidth,height=0.75\linewidth]{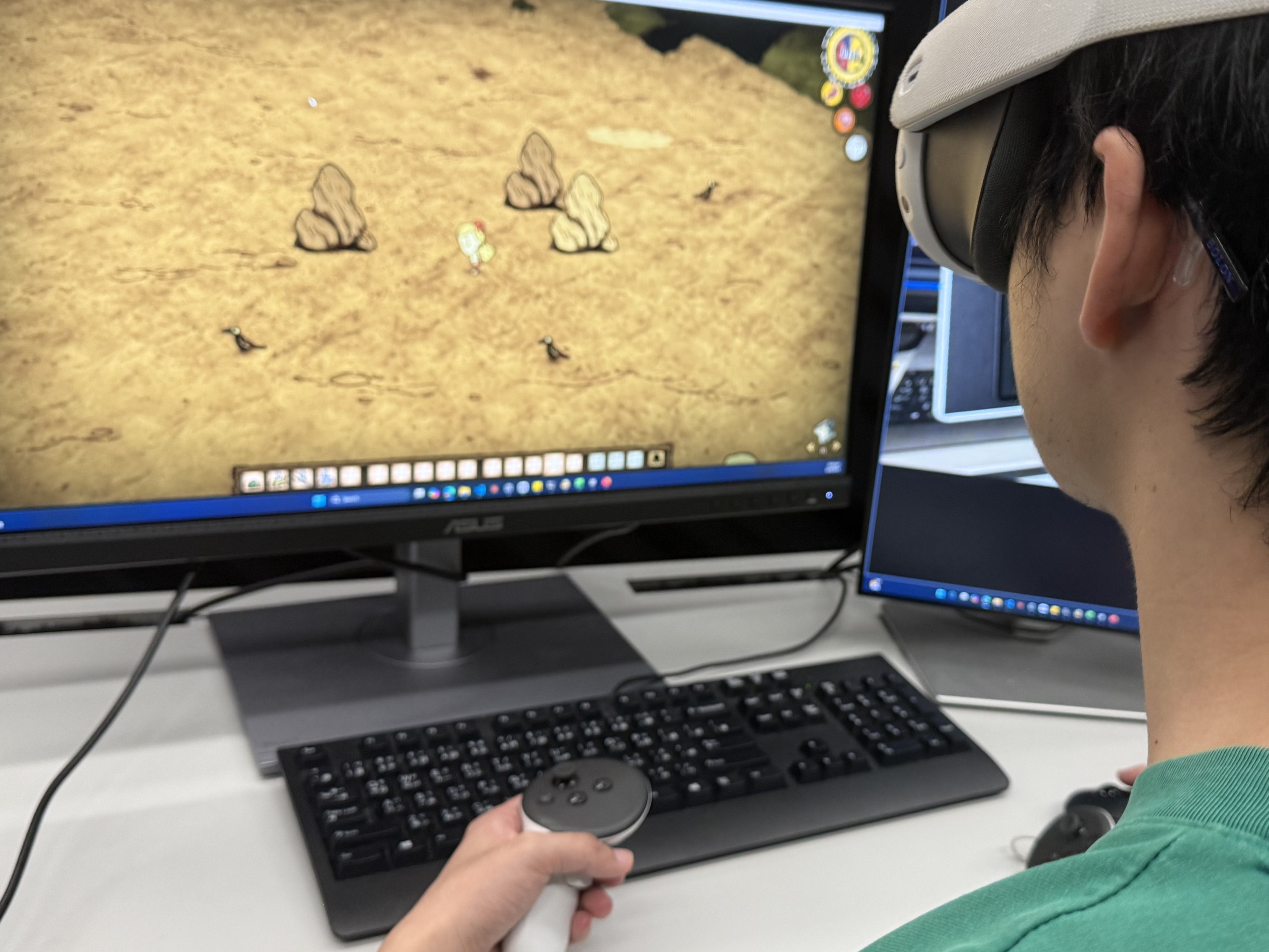}
    \caption{Gameplay}
  \end{subfigure}
  \caption{Formative-study tasks: (a) painting on a tablet, (b) preparing a drink, and (c) playing a survival game.}\label{fig:pilot-tasks}
  \Description{Three panels arranged in a single row show, from left to right, a headset-wearing participant painting on a tablet, preparing a drink, and playing a survival game.}
\end{figure}

We selected three tasks spanning physical manipulation, GUI interaction, and controlled drawing (Figure~\ref{fig:pilot-tasks}). Each comprised six stages with one to four steps per stage. Participants used identical materials and goals in both conditions.

\textbf{Cocktail preparation} involved measuring ingredients, following an addition sequence, and shaking. \textbf{Gameplay} combined physical inputs and GUI decisions in a survival game. \textbf{Digital painting} involved selecting tools and controlling strokes to draw an apple. These tasks offered different opportunities to inspect actions, anticipate consequences, and compare methods. Each participant completed all six task--condition combinations, with starting task and condition orders counterbalanced across participants.

\subsubsection{Method.}
The study comprised a task session (90--120 minutes) and a questionnaire and interview session (approximately 30 minutes). After familiarization, each participant completed all three tasks in both conditions, for six executions. Starting task and condition orders were balanced, with rest breaks between executions.

Participants thought aloud during execution. After each generative-condition task, the experimenter asked them to inspect any available panels they had not viewed. We distinguished this prompted inspection from voluntary use during execution and required branch choices.

In the second session, participants completed SUS~\cite{sus_Brooke_96}, NASA-TLX~\cite{nasatlx1988}, and component rankings, then discussed usefulness, difficulties, and generation delays. We organized think-aloud and interview accounts by candidate, contextualized by questionnaires. As prototypes evolved across participants, we used these accounts to guide iterative refinement.

\subsubsection{Results.}\label{sec:formative-results}

Participants recognized the potential value of additional previews but often prioritized completing the current action over exploring further. The findings below show how relevance, access effort, and representation shaped their engagement.

\paragraph{Next-Step Preview: Immediate Action Support Remained the Priority.}
All six valued detailed next-step support and considered the guidance sufficient, despite differing media preferences. Exploration competed with execution for attention, motivating ordinary next-step guidance as the foundation and separate examination of representation choice.

\paragraph{Deviation\textendash{}Consequence Preview: Visible Consequences Were Consulted, Further Exploration Was Rare.}
Access shaped participants' engagement with Wrong Case. Once the panel was directly visible, P5 and P6 reported consulting its information and checking before acting whether they might make the depicted mistake. The panel remained clickable for further exploration, replacing the separate access button. Voluntary use of the exploration entry was sparse: P2 opened it once at the start of a task, while P1, P3, and P4 did not activate it during execution.

Participants valued the concept but felt little need to investigate unmade errors while attending to the task. After further inspection of the panel, all six preferred information about errors relevant to their own situation and how to correct them over additional videos of erroneous actions. These accounts distinguish pre-action consequence inspection from post-error correction.

\paragraph{Alternative-Method Preview: Choices Became Meaningful When Connected to the Current Decision.}
P2, P3, P5, and P6 acknowledged the potential usefulness of alternative methods. P2 and P3 did not voluntarily open the optional alternative-method control during execution; inspection followed post-task prompting. P3 explained, \emph{``I think it is useful, but I do not want to think about these things while carrying out the task.''} P3 anticipated considering alternatives during a second attempt or after gaining some understanding of the procedure. P4 prioritized the next action, expecting the same immediate result without inspecting alternatives.

In V3, the required Alternative-Method Preview panels connected methods to their outcomes and affected steps. P6 commented, \emph{``I like these choices. They explain what I am doing now and make me feel in control of the task.''} These accounts suggest that perceived value depended on the choice's relevance to the current task.
\paragraph{Recovery: Demand for Reliable Post-Error Assistance.}
All six valued recovery support, typically seeking it after mistakes, but incorrect guidance discouraged continued use. After encountering an initial failure in which the VLM decomposed recovery into inappropriate actions, P1-P3 stopped invoking the recovery control and preferred to explore using other guidance they considered reliable. As P1 put it, \emph{``Exploring on my own is faster and more accurate than waiting for guidance.''} P1 also noted that feedback from the task software or environment already supported trial and error. These accounts emphasized the need for reliable correction that justified waiting.

\paragraph{Multi-Step Preview: Progress Orientation Was Valued More Than Extended Look-Ahead.}
All six participants said Multi-Step Preview gave them a sense of control over task progress. Yet its use was concentrated near the beginning: P1-P3 checked Multi-Step Preview at task onset, and no participant opened its exploration control midway through execution. P4 explained that checking additional previews several stages ahead would be distracting, especially while still learning the procedure. Its clearest perceived value was orientation, not extended look-ahead.

\paragraph{Action-Variation Preview: Interest in Experimentation Was Deferred Until After Familiarization.}
Action-Variation Preview allowed users to adjust predefined parameters or targets of the current action and inspect the resulting video. P1-P3 found the feature interesting after viewing it, but did not spontaneously invoke it during task execution. P4 found its additional controls particularly burdensome: an already crowded interface led to yet more interaction after opening the panel. P2 anticipated using the feature only with sufficient prior knowledge and a strong desire to understand the task, while P1 felt that simple hints followed by self-directed exploration could be faster.

P3 expressed the distance between this exploratory activity and the immediate goal: \emph{``Before I finish the task, I do not want to explore questions beyond the next-step guidance.''} Exploring changes to the current action could still feel removed from completing it, suggesting a role in later practice or deliberate exploration.

\paragraph{Representation Choice: Users Weighed Action Clarity Against the Cost of Inspection.}\label{sec:formative-media}
Participant accounts described AR cues and images as a lightweight starting point, with important task-specific exceptions. They turned to video when these representations left the next action unclear or when they were uncertain what their execution should produce. Participants valued prerecorded clips' perceived accuracy, but clips covering more than the needed action required searching for the relevant moment, operation, or target, making details easier to miss. In time-sensitive gameplay, participants valued AR cues for reducing the effort of matching depicted objects and locations to the live scene. No participant paused ongoing gameplay to request or watch a video.

iPad painting provided an important contrast. Generated videos attracted the attention of P2, P3, P5, and P6, and all six described relying primarily on video because the AR stroke animations were difficult to interpret. This preference persisted even though AR and video depicted the same motion trajectory and AR guidance was available sooner. In this setting, seeing the movement within the video made the action easier to read. Occasional generation errors and fast playback frustrated P1 and P4 and interrupted their task execution. Outside painting, video was generally consulted when an operation was complex enough to warrant waiting.

\subsection{Interface Design Rationale}\label{sec:design-rationale}

\subsubsection{Design Considerations}\label{sec:formative-design-considerations}
The formative findings informed three design considerations.

\textit{DP1. Make consequences directly accessible.} Later participants consulted visible consequence information, while optional exploration saw little use. Consequence cues and images should therefore be available alongside the current instruction without requiring additional inspection controls.

\textit{DP2. Tie method comparison to actionable decisions.} Later accounts valued choices connected to task continuation. Method comparison should therefore connect consequences to a decision the user can act on.

\textit{DP3. Match representation and timing to the action.} Task-dependent media preferences and reported playback and generation problems highlighted the importance of action clarity and timely delivery (Section~\ref{sec:formative-media}). Representation choice should follow the information needed at the current action while accounting for waiting and inspection costs.

\subsubsection{Final Interface}\label{sec:final-interface-scope}
\begin{figure}[htbp]
  \centering
  \begin{subfigure}[t]{0.5\linewidth}
    \centering
    \includegraphics[width=\linewidth,height=27mm,keepaspectratio]{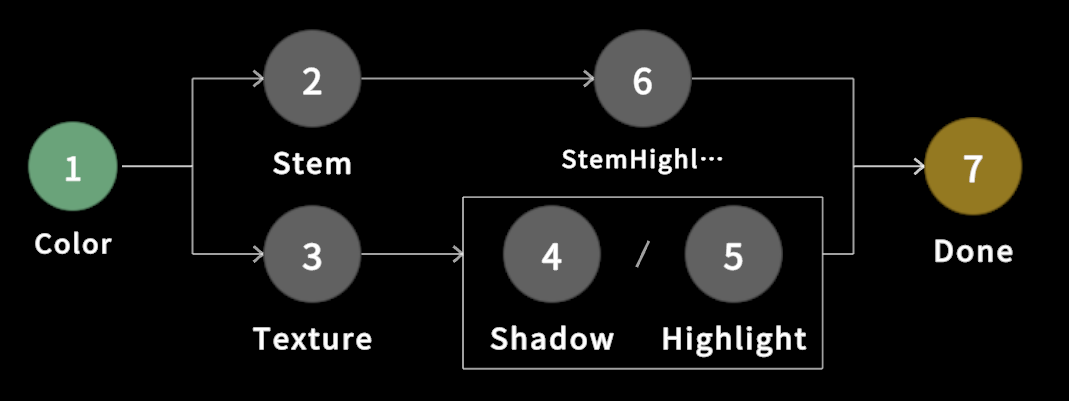}
    \caption{Multi-Step Preview Panel.}\label{fig:final-map}
  \end{subfigure}
  \hfill
  \begin{subfigure}[t]{0.235\linewidth}
    \centering
    \includegraphics[width=\linewidth,height=27mm]{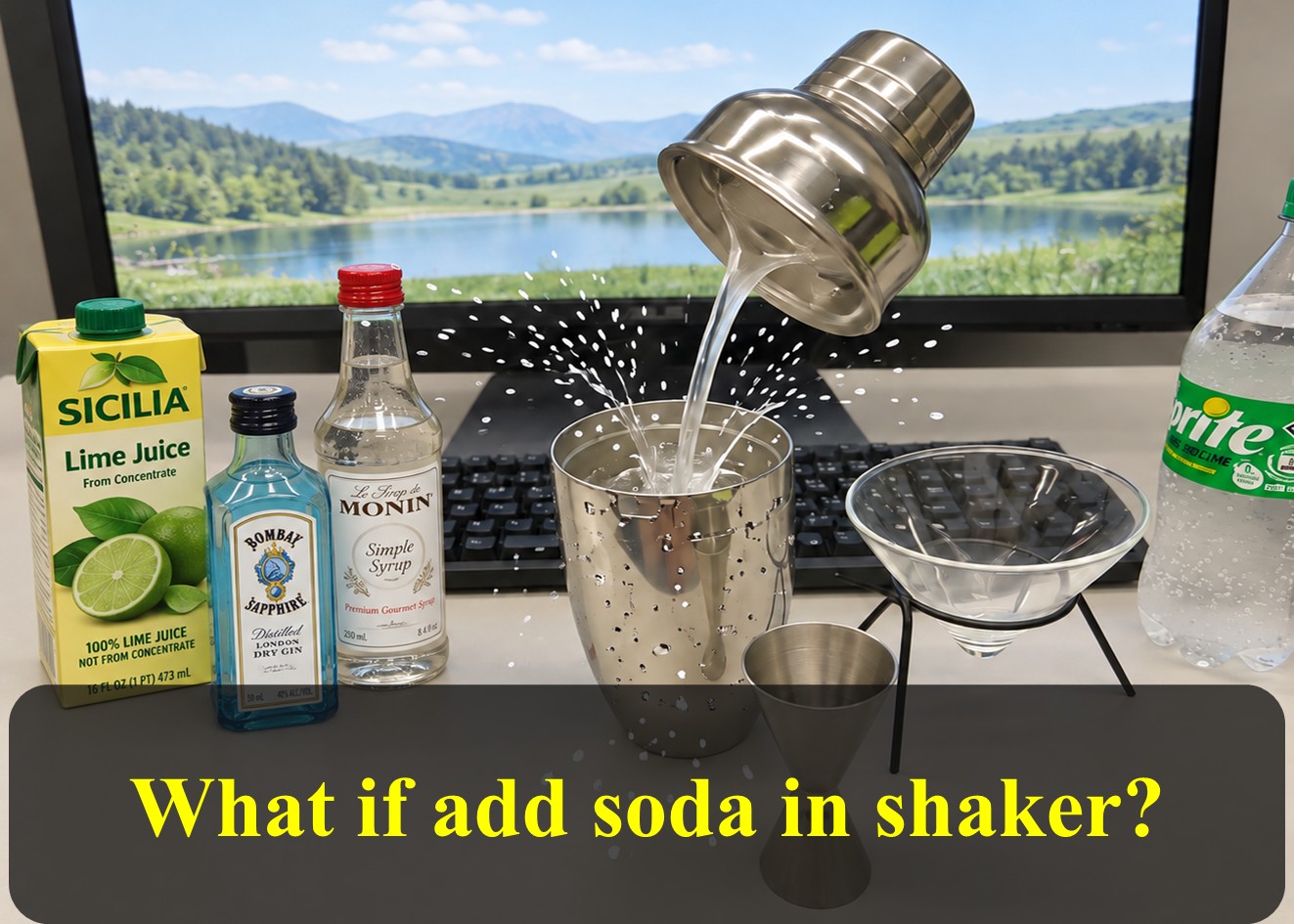}
    \caption{Deviation-Consequence Preview Panel.}\label{fig:final-wrong-case}
  \end{subfigure}
  \hfill
  \begin{subfigure}[t]{0.235\linewidth}
    \centering
    \includegraphics[width=\linewidth,height=27mm]{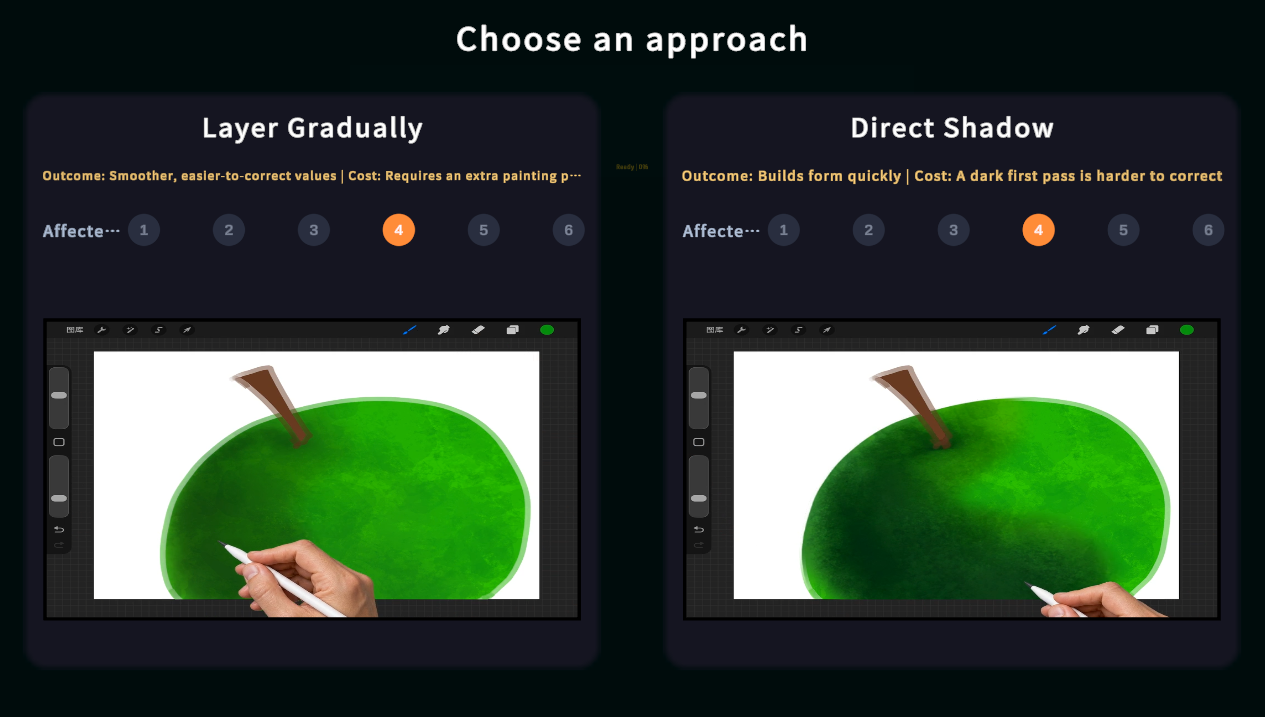}
    \caption{Alternative-Method Preview Panel.}\label{fig:final-alternative}
  \end{subfigure}
  \caption{\textbf{Integrated guidance and independent preview components.} (a) Multi-Step Preview Panel provides clickable access to generated and upcoming previews while keeping the execution step unchanged; (b) Deviation-Consequence Preview pairs a soda-in-shaker cue with an adverse-outcome image; (c) Alternative-Method Preview Panel compares painting approaches. Next-Step Preview provides ordinary next-step guidance; Multi-Step Preview and the two comparison/consequence panels are independent components.}\label{fig:final-panels}
  \Description{Three panels show a clickable Multi-Step Preview panel, a Deviation-Consequence Preview pairing a soda-in-shaker cue with an adverse-outcome image, and an Alternative-Method Preview comparing two painting approaches.}
\end{figure}

Beyond ordinary next-step guidance, Foresee retains Deviation\textendash{}Consequence, Alternative-Method, and Multi-Step Preview (Figure~\ref{fig:final-panels}).

\textbf{Deviation\textendash{}Consequence Preview} is a visible panel pairing a brief ``What if~\ldots?'' cue with a wrong-outcome image, making a hypothetical deviation and its consequence inspectable before action.

\textbf{Alternative-Method Preview} compares methods at selected branches. It compares feasible methods through their outcomes, costs, affected steps, and action demonstrations. From V3 onward, each task has two fixed decision points; the user must select a method before guidance advances. Other preview inspection remains optional.

\textbf{Multi-Step Preview} is a compact, clickable step overview. It provides task-progress orientation and access to previously generated previews and previews of upcoming steps. Selecting a node changes the viewed content without changing the current execution step.

We retain Multi-Step Preview because its visible overview supports orientation without requiring exploration, while optional node access adds depth. Action-Variation requires additional manipulation and is set aside from the main workflow. Recovery addresses observed errors and falls outside our prospective-guidance scope. Here, \emph{minimal} denotes a compact default presentation with fewer standalone exploration controls.

\section{System Design and Implementation}\label{sec:generation-rationale}

\begin{figure}[htbp]
  \centering
  \includegraphics[width=\linewidth]{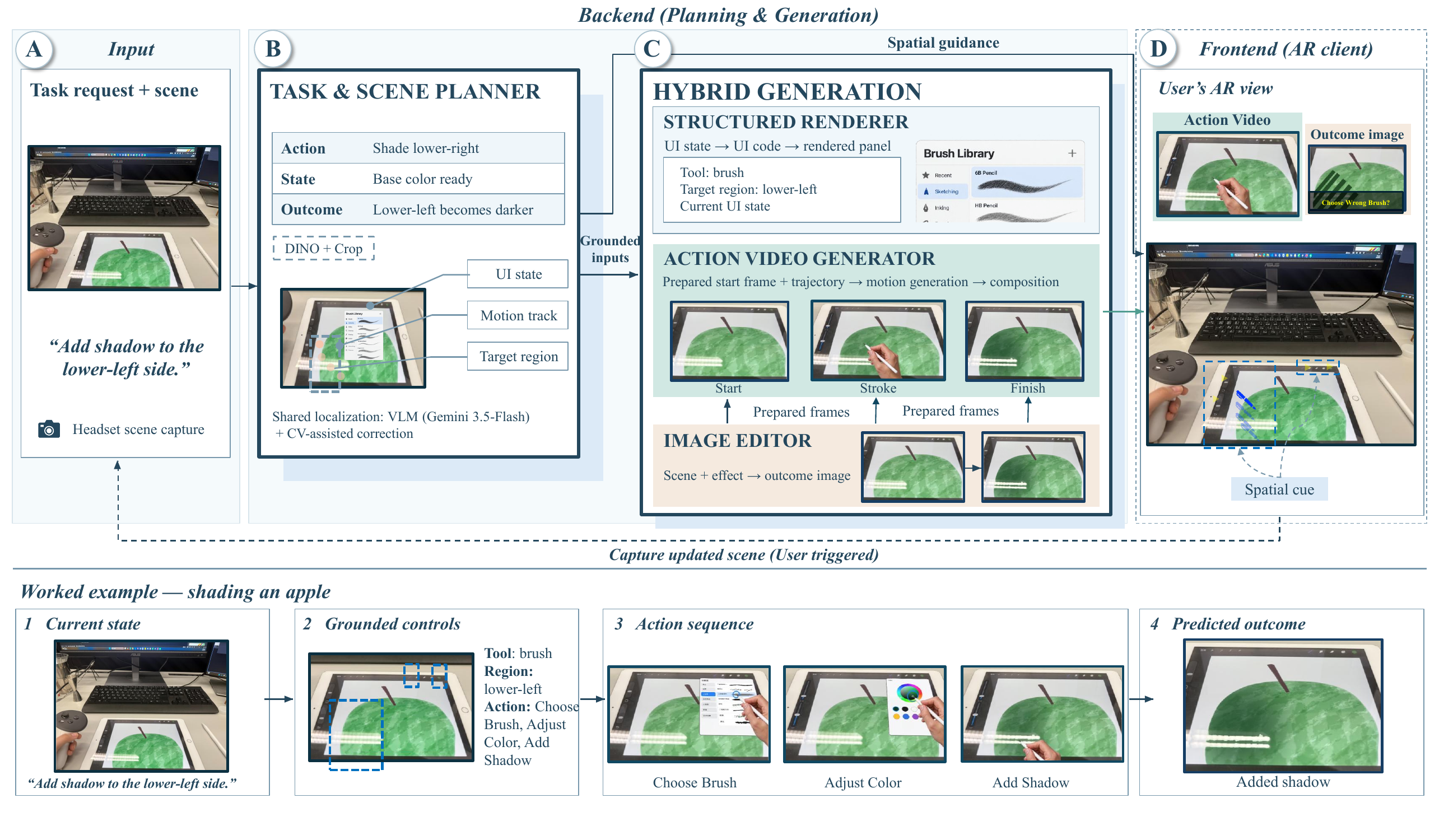}
  \caption{\textbf{System architecture and generation workflow.} (A) The AR client supplies the task request and current scene. (B) An apple-shading example traces task decomposition, scene grounding, generation-input preparation, and conditioned generation. (C) Hybrid generation combines structured UI rendering, action-video generation, and image editing. (D) Spatial guidance, action videos, and outcome images return to the AR client.}
  \Description{A task request and captured scene enter the task and scene planner, which supplies grounded inputs to structured rendering, action-video generation, and image editing. The AR client on the right receives spatial cues, action videos, and outcome images. User-triggered scene capture supplies updated input. A lower apple-shading example traces the current scene, grounded controls, brush selection, color adjustment, shading, and resulting image.}
  \label{fig:method}
\end{figure}

Foresee implements the interface developed through the formative study (Section~\ref{sec:final-interface-scope}) through reviewed plans and scene-conditioned generation. To support valid, interpretable guidance with acceptable waiting, it combines structured rendering for rule-defined content with diffusion for complex appearance and motion. Figure~\ref{fig:method} illustrates the workflow through apple shading.

\subsection{Overview}

Foresee uses Qwen3.8\mbox{-}27B to organize task references into a predefined plan of stages, atomic actions, error-sensitive steps, and method choices, reviewed before deployment~\verifiedcite{qwen38model}. During use, it combines the selected step with the current scene image to prepare action guidance and consequence previews.

The hybrid pipeline prepares video inputs using structured rendering and compositing for rule-defined GUI content, or generated outcome images when needed. These frames guide action-video generation. The client displays available images and videos alongside AR cues, allowing users to inspect the guidance while continuing the task.

\subsection{Procedural Representation and Preview Generation}

Each action specifies a precondition, intended state transition, observable outcome, and relationship to later steps. Separating actions from resulting states supports complementary representations: Deviation\textendash{}Consequence Preview depicts a specified hypothetical outcome, while next-step and alternative-method videos demonstrate execution. Multi-Step Preview organizes access to guidance across the planned procedure.

Foresee's hybrid pipeline combines structured rendering with image and video generation. For rule-defined GUI content, the planner produces a schema-validated rendering description that is compiled into rendering operations for text, icons, panels, and layout. Image editing constructs an action-relevant starting image or a hypothetical outcome; video generation supplies hand, tool, and other action motion. Foresee uses Qwen-Image-Edit-2511 with a four-step Lightning adapter for image editing and HunyuanVideo~1.5 for action demonstrations~\verifiedcite{qwenimageedit2511,qwenimagelightning,hunyuanvideo15}. This division controls interface structure while generating detail and motion beyond the rendering rules.

In the apple-shading example (Figure~\ref{fig:method}), the input pairs the canvas screenshot with an instruction to shade the apple's lower-right side. The specification identifies the target region, motion, and intended shaded state. Alongside the instruction, spatial cues, an outcome image, and an action video show where to act, how, and with what result.

\subsection{Runtime Interaction and Delivery}

Generation timing depends on each step's media requirements. When a video requires a generated end frame, image generation precedes video generation, increasing the wait. When code-rendered content and compositing suffice, this preparation is shorter. Available images can be displayed while video generation continues and remain accessible if it fails, allowing participants to retry.

Next-step generation takes priority over Deviation\textendash{}Consequence (Wrong Case) image generation. Wrong Case waits when next-step processing occupies the required generation resources; otherwise, it can run asynchronously alongside next-step processing.

Multi-Step Preview retrieves available content when a node is selected, allowing inspection while keeping the current execution step unchanged. Generation for a later step uses that step's planned action and the current screenshot rather than a simulated intermediate state. It therefore previews planned guidance in the currently observed scene.

At designated error-sensitive steps, Deviation\textendash{}Consequence Preview presents its cue and outcome image alongside the current instruction. At each task's two fixed branch points, Alternative-Method Preview presents the available methods and their consequences; participants must select a method before guidance advances. This requirement gates guidance progression, not their ability to act in the task environment. Method videos play automatically when available, while other preview inspection remains optional. Step progression is triggered by task state in gameplay and advanced manually in painting and cocktail preparation.

The client associates outputs with actions, and the service caches intermediate content within each session. Unfinished jobs continue after step advancement, with no separate cancellation or stale-output handling; late completion does not block progression.

\subsection{Stability Evaluation}\label{sec:technical-feasibility}

\begin{table}[htbp]
  \centering
  \small
  \setlength{\tabcolsep}{4pt}
  \caption{Historical service-request outcomes and server processing times (seconds). Timing summaries use 477 completed image requests and 694 completed video requests with valid timing and no recorded communication or upstream exceptions; 11 completed image requests lacked timing. $M$ is the mean, Median the 50th percentile, and P95 the 95th percentile. Times cover server processing, excluding full client delivery; video continuations exclude the preceding image stage.}
  \Description{Historical service-request outcomes and server processing times. Image generation has 523 requests, 27 failures, and 496 completions; mean, median, and 95th-percentile processing times are 13.63, 10.11, and 35.90 seconds. Video generation has 712 requests, 14 failures, and 698 completions; corresponding times are 19.74, 10.16, and 61.13 seconds. Timing summaries use 477 completed image requests and 694 completed video requests with valid timing and no recorded communication or upstream exceptions; 11 completed image requests lack timing. Times exclude full client delivery, and video continuation times exclude the preceding image stage.}
  \label{tab:runtime-performance}
  \begin{tabular}{lrrrrrr}
    \toprule
    Type & Requests & Failed & $M$ (s) & Median (s) & P95 (s) & Completed \\
    \midrule
    Image & 523 & 27 & 13.63 & 10.11 & 35.90 & 496 \\
    Video & 712 & 14 & 19.74 & 10.16 & 61.13 & 698 \\
    \bottomrule
  \end{tabular}
\end{table}

We analyzed historical service logs covering the user-study period and development tests. A single user request may trigger multiple server requests, so these service-level records provide a finer-grained view of runtime reliability across intermediate processing stages. Of 1,237 image/video requests, we excluded two final failures caused by upstream unavailability or connection refusal from both numerator and denominator. The retained 1,235 requests included 1,194 completions and 41 final failures (3.32\%; Table~\ref{tab:runtime-performance}). Failures comprised target/action recognition (23), invalid coordinates (8), screen registration (4), local generation timeouts (4), and HTTP 400 parameter errors (2). Completion denotes a finished service job. We also examined generated-video quality qualitatively through user-study observations and interviews (Section~\ref{sec:qual-operational-information}).

\section{User Study}

We conducted a within-subject study ($N=24$) to examine preview use, interpretation, and perceived usefulness (RQ1), and compare task execution and post-task understanding (RQ2). The study compared AR guidance, AR with generated action video (Video Gen), and the full Foresee interface (Preview). Each participant experienced all three conditions on different tasks.

\subsection{Tasks and Conditions}

We used the same three task domains as the formative study (Section~\ref{sec:formative-pilot}): Digital Painting, Gameplaying, and Cocktail Making. These domains provided contrasting combinations of device, task content, temporal demands, and action type. Each task contained error-sensitive steps and two branch points with different intermediate consequences but the same final task goal.

Step-level success required the instructed effect and target state, not matching the demonstrated movement or route. Alternative methods achieving the required effect counted as successful. In Digital Painting, this meant producing the required component and painting effect. In Gameplaying, each step had to achieve its intermediate goal, with overall completion defined by constructing the Science Machine. In Cocktail Making, actions had to satisfy the specified ingredient and preparation requirements. Omissions indicated unachieved step effects, not valid alternative routes. Scoring thus distinguished final-goal completion from intermediate requirements.

All three conditions shared ordinary next-step instructions, reference images, and situated AR action cues. The conditions incrementally added generated video and the future interface:
\begin{itemize}[leftmargin=*]
    \item \textbf{AR (A):} the shared next-step guidance, including reference images and AR cues.
    \item \textbf{Video Gen (B):} the shared next-step guidance plus generated action video.
    \item \textbf{Preview (C):} the Video Gen condition plus Foresee's future interface described in Section~\ref{sec:final-interface-scope}.
\end{itemize}

Task goals, success criteria, and the underlying step-advancement mechanism were shared across conditions: task state triggered advancement in gameplay, while participants advanced manually in the other tasks. Preview additionally required a method selection at its two fixed decision points before guidance advanced. The two video-supported conditions shared playback controls. Generated previews used the current scene image and a predefined step prompt that a language model could adapt; failed generation attempts were retried.

Each participant performed each task and condition once across three trials. Task--condition assignments and task order were counterbalanced, with eight participants per combination. Within-task condition comparisons therefore involved different participants.

Across conditions, the experimenter provided backend GUI-target assistance when participants needed help locating a target in gameplay and painting, without spoken or pointing assistance. Cocktail preparation used no such assistance.

\subsection{Participants and Apparatus}
We recruited 24 participants (U1--U24; 13 male, 11 female; age: $M=26.7$, $SD=6.5$ years), of whom 13 had prior VR/AR experience. Participants rated their prior familiarity with each task domain on a scale from 1 to 7. For cocktail making, gameplay, and digital painting, respectively, 4, 7, and 8 participants selected ratings of 4--6, while 11, 11, and 3 selected a rating of 1. All participants used an Oculus Quest 3 connected to a generation service running on a server with one NVIDIA H100 GPU\@.

\subsection{Measures}
\paragraph{Task execution.}
We recorded whether each step was completed correctly, omitted, or performed incorrectly, including whether an error was subsequently corrected. Pre-action preview use was recorded separately from execution outcomes. Observations were checked with participants afterward, distinguishing corrected execution from correct execution without a recorded error. We report execution categories that distinguish corrected errors, unresolved errors, and correct execution without a recorded error.

\paragraph{Post-task understanding.}
For each task, we designed a quiz assessing factual recall, basic conceptual understanding, and knowledge transfer, following prior work~\cite{zhang2025reflective}. Most quiz items addressed task knowledge that participants could also acquire through interaction with the task environment and exploration, rather than knowledge available exclusively in the additional preview panels. The Preview condition added step-specific explanations of selected action--consequence relationships, while participants across all conditions could also draw on ordinary guidance and task experience. Before task execution, participants identified items whose answers they already knew. We excluded these participant-specific items from both the numerator and denominator, calculating post-task accuracy over the remaining eligible items. Eligible item counts and respondent sample sizes are reported with the quiz results in Section~\ref{sec:quant-results}.

\paragraph{Subjective experience.}
We collected seven-point adaptations of the System Usability Scale (SUS)~\cite{sus_Brooke_96} and NASA Task Load Index (NASA-TLX)~\cite{nasatlx1988}. To compare representations of ordinary next-step guidance, participants ranked AR cues, ordinary reference images, and generated action video from 1 (most preferred) to 3 (least preferred) for each task. Here, images refer to the ordinary reference images shared across conditions, distinct from the additional consequence previews. A separate overall ranking ordered five components: Wrong Case, Multi-Step Preview, Alternative-Method Preview, Video, and AR, with lower ranks indicating stronger preference. Seven-point ratings assessed the choice interface's alignment with personal priorities, perceived control, and interruption. Think-aloud accounts and semi-structured interviews complemented these measures.

\paragraph{Questionnaire scoring.}
Higher SUS scores indicate greater usability, while higher NASA-TLX scores indicate greater workload.

Condition comparisons focused on execution, understanding, and experience. Total task time combined execution, variable generation waits, and think-aloud activity, so completion time was not compared across conditions.

\subsection{Procedure}
The study comprised a task session (40--60 minutes) and a questionnaire, quiz, and interview session (approximately 30 minutes). After briefing and informed consent, participants had 10 minutes of system familiarization, then completed three assigned trials with five-minute breaks and think-aloud invitations. Questionnaires and task quizzes followed all trials, then a semi-structured interview covered overall experience, usefulness, difficulties, and generation delays. Participants did not consume the alcoholic drinks they prepared. Headset discomfort prompted a pause and rest.

\begin{figure*}[htbp]
  \centering
  \includegraphics[width=\linewidth,height=0.75\textheight,keepaspectratio]{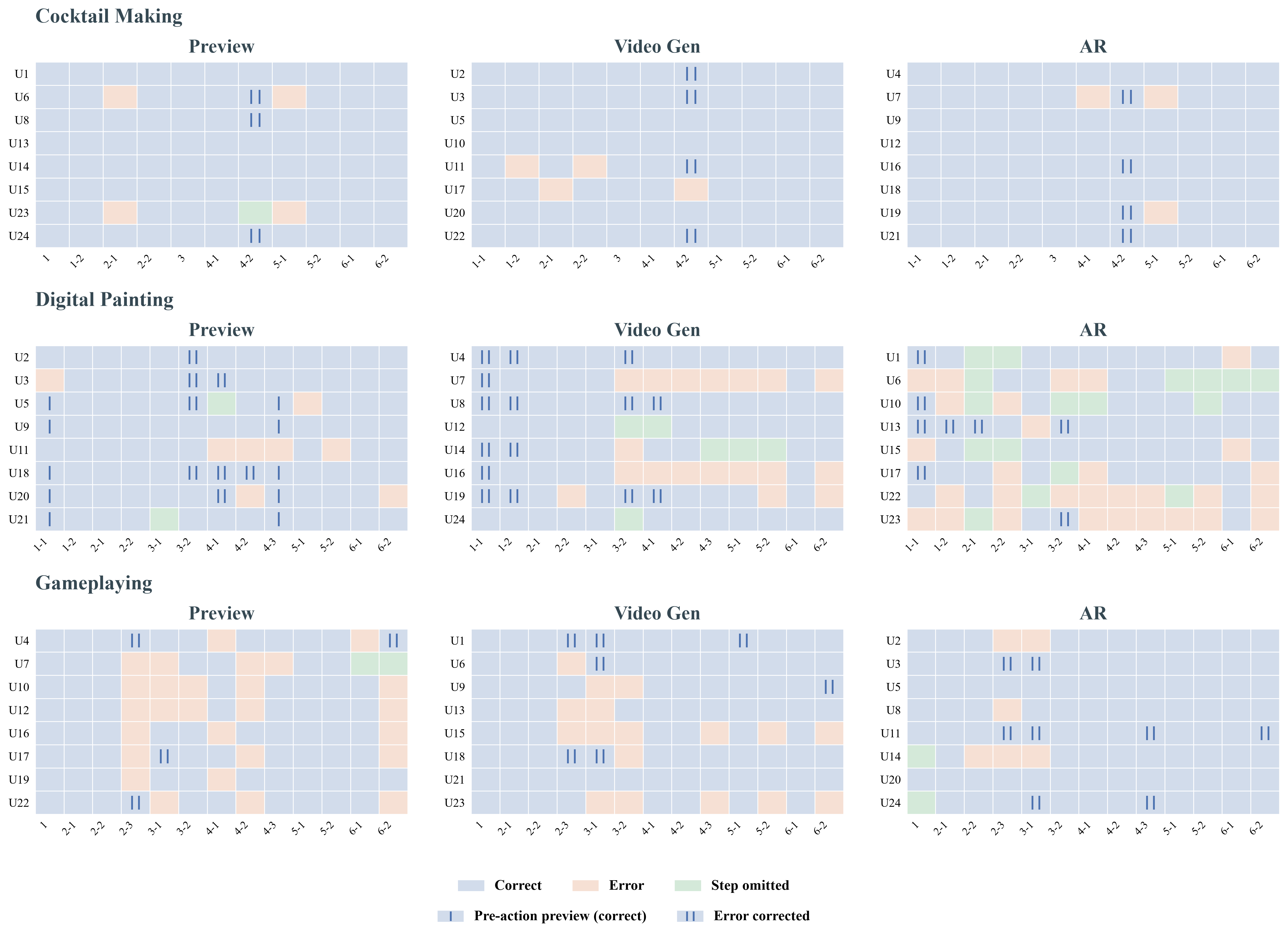}
  \caption{Step-level execution annotations across all three tasks. Task blocks show Cocktail Making, Digital Painting, and Gameplaying; columns of panels show Preview (C), Video Gen (B), and AR (A). Each panel contains eight participants, with 11 annotated steps in cocktail making and 13 in each other task. Blue cells indicate correct execution, peach cells errors, and green cells omissions. Single vertical marks identify correct action following a pre-action preview; double marks identify corrected errors.}
  \Description{Nine heatmaps arrange three tasks by three guidance conditions. Rows within each panel identify participants and columns identify steps. Cell colors distinguish correct execution, errors, and omissions, while single and double vertical marks distinguish pre-action preview followed by success from error correction.}
  \label{fig:all-tasks-step-annotations}
\end{figure*}

\subsection{Quantitative Results}\label{sec:quant-results}

Results are task-specific because action requirements, content, and inspection opportunities differ. Pooled execution and quiz comparisons detected no significant condition differences; the analyses below examine differences within each task. Within-task comparisons used participant-level summaries from independent groups. Execution comparisons used two-sided exact permutation tests, and quiz comparisons used Welch's $t$ tests with an exact-permutation sensitivity analysis. For painting, we report mean differences ($\Delta$) and pooled-SD Cohen\textquotesingle{}s $d$, with positive values favoring the first-named condition. The $d$ estimates are approximate because they use rounded summaries and eight participants per condition. SDs and $\Delta$ for percentage outcomes are expressed in percentage points. Task-specific analyses are exploratory, with Holm correction families and sensitivity results specified below.

\subsubsection{\textbf{Execution Accuracy. }}
For Digital Painting, we calculated three rates for each participant over 13 annotated steps: correct completion without a recorded error, any recorded error, and unresolved error. Omissions remained in the denominator. An error followed by correction counted toward the any-error rate, but not the unresolved-error rate or error-free-correct rate.

Correct completion without a recorded error was higher with Preview ($M=82.7\%$, $SD=10.7$) than AR ($M=47.1\%$, $SD=21.5$; $\Delta=35.6$, 95\% CI [20.2, 51.0], $d\approx2.10$, $p_{\mathrm{adj}}=.0042$). Video Gen scored $M=62.5\%$ ($SD=21.1$). Neither Video Gen--AR ($d\approx0.72$, $p_{\mathrm{adj}}=.794$) nor Preview--Video Gen ($d\approx1.21$, $p_{\mathrm{adj}}=.299$) was significant. Holm correction covered nine painting comparisons: three contrasts for each of the three rates.

Any-error rates were 15.4\%, 31.7\%, and 36.5\% for Preview, Video Gen, and AR; unresolved-error rates were 7.7\%, 17.3\%, and 28.8\%. None of these separate contrasts survived correction. Correct completion without a recorded error thus provides the clearest separation between conditions, accounting for both incorrect execution and omitted requirements.

Figure~\ref{fig:all-tasks-step-annotations} shows execution histories across all three tasks, including 11 corrected errors in Cocktail Making, 31 in Digital Painting, and 19 in Gameplaying. In Digital Painting, ten step records across five Preview participants indicate correct action following a pre-action wrong-case cue. These records situate preview inspection within execution, distinguishing guidance preceding a successful action from correction after an error.

Some participants reported inferring cocktail actions from instructions and common sense, while gameplay supplied state-triggered AR guidance in every condition. These resources contextualize preview use but do not establish why condition comparisons were nonsignificant.

\begin{figure*}[htbp]
  \centering
  \begin{subfigure}[t]{0.49\linewidth}
    \centering
    \includegraphics[width=\linewidth]{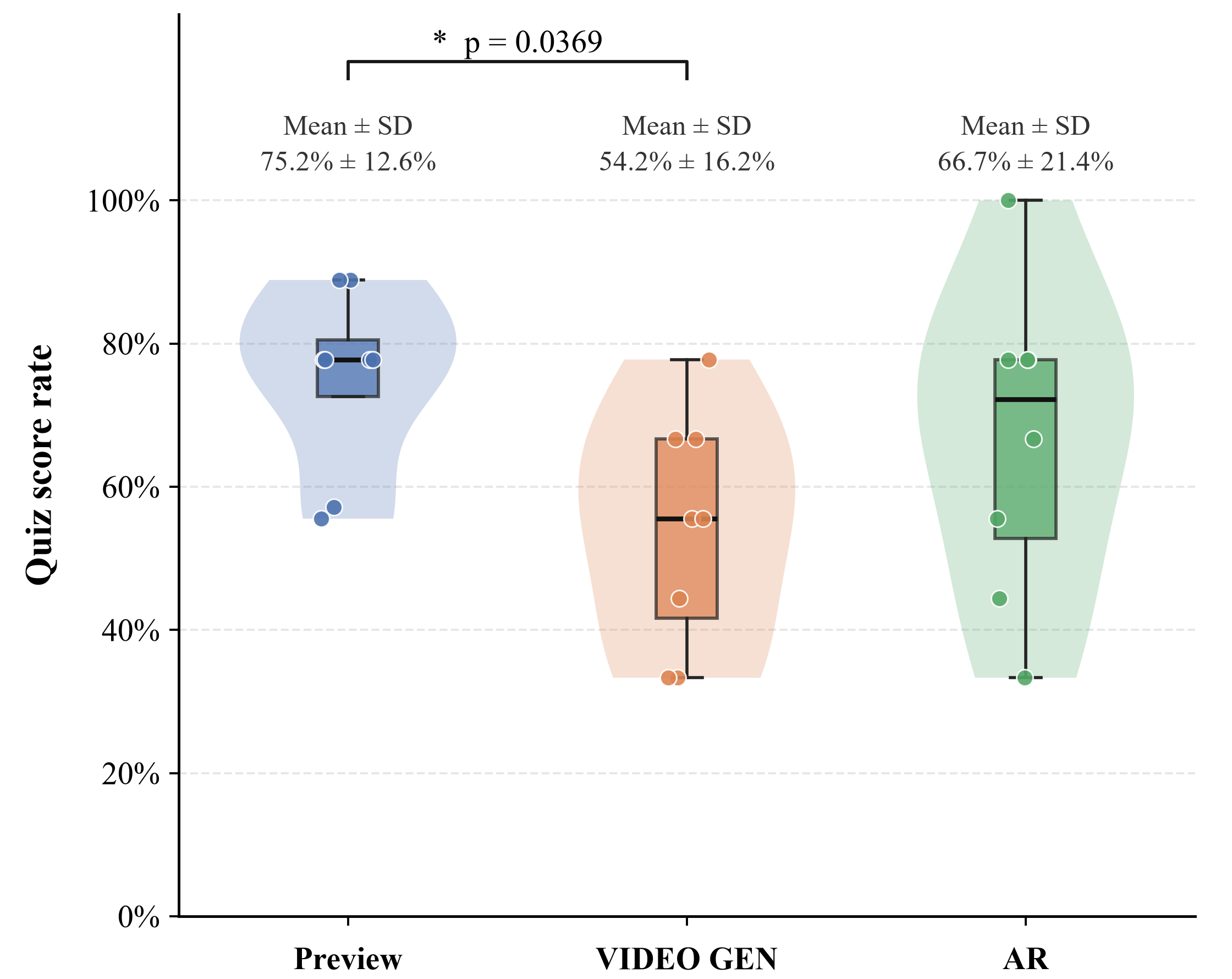}
    \caption{Post-task quiz accuracy in Digital Painting task.}
    \label{fig:painting-quiz}
  \end{subfigure}\hfill
  \begin{subfigure}[t]{0.49\linewidth}
    \centering
    \includegraphics[width=\linewidth]{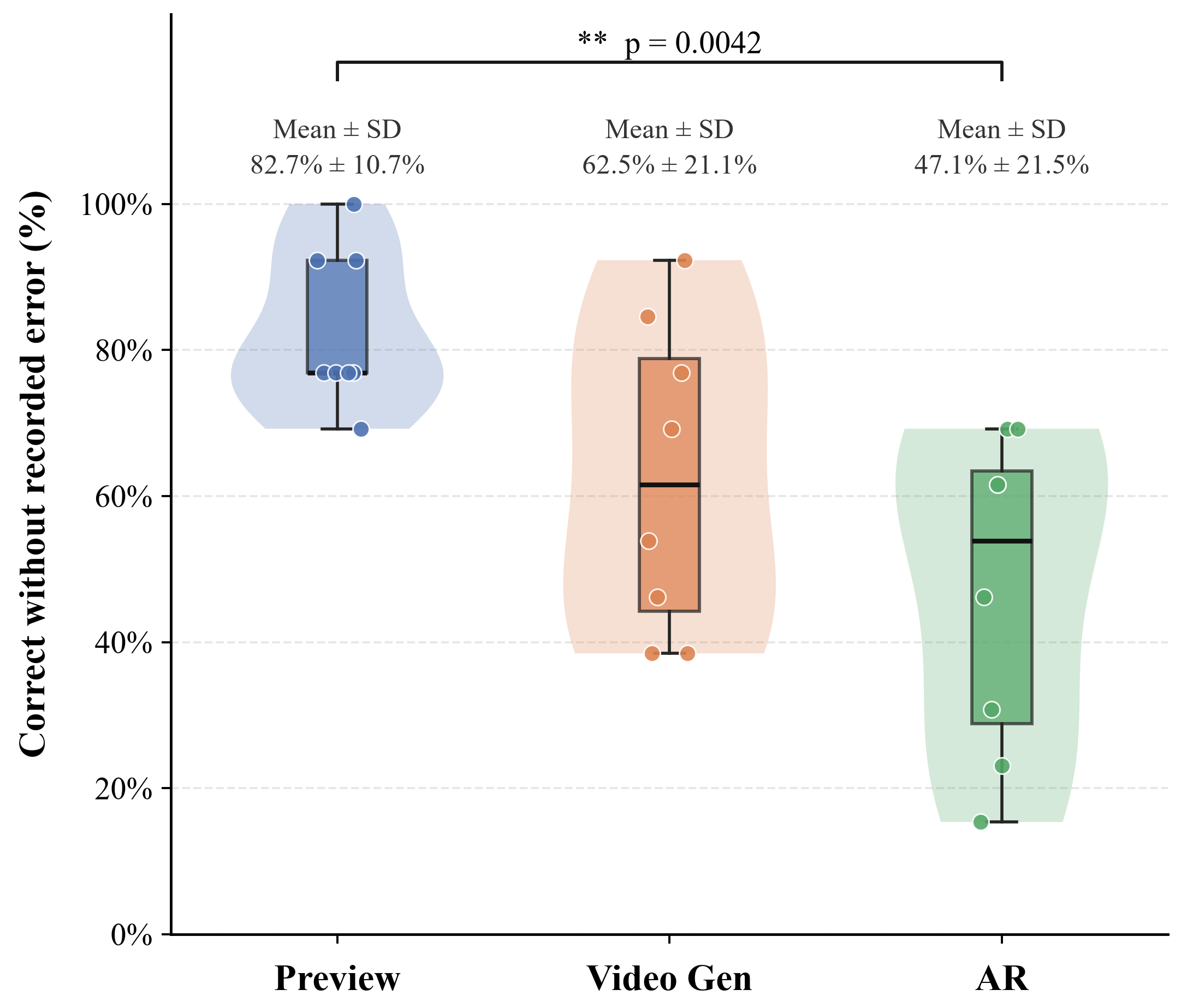}
    \caption{Correct execution without a recorded error in Digital Painting task.}
    \label{fig:painting-error-free}
  \end{subfigure}
  \caption{Understanding and execution outcomes in Digital Painting. Preview, Video Gen, and AR correspond to C, B, and A. Dots show individual participant scores; labels report means and standard deviations. (a) Quiz accuracy was higher for C than B (task-wise Holm-adjusted $p=.0369$). (b) Correct-without-recorded-error rates use all 13 annotated steps per participant, including omissions, with eight participants per condition; C exceeded A (Holm-adjusted $p=.0042$).}
  \Description{Two panels combine violin plots, boxplots, and participant scores for Digital Painting. The left panel shows quiz means of 75.2, 54.2, and 66.7 percent for Preview, Video Gen, and AR, with a Preview versus Video Gen annotation of task-wise adjusted p equals 0.0369. The right panel shows correct-without-recorded-error means of 82.7, 62.5, and 47.1 percent, with a Preview versus AR annotation of adjusted p equals 0.0042.}
  \label{fig:painting-understanding-execution}
\end{figure*}

\subsubsection{\textbf{Quiz Scores.}}

Quiz accuracy was calculated per participant over eligible items, then averaged within each condition. Scoring covered 665 eligible item responses across 71 participant--task scores. In Preview/Video Gen/AR order, eligible items per participant were 10, 10, and 4--10 for Cocktail Making; 7--9, 6--9, and 9 for Digital Painting; and 9--10, 4--10, and 10 for Gameplaying. Each task--condition cell retained eight respondents, except Cocktail Making--Video Gen ($n=7$), where one participant had no eligible items. Cocktail Making--AR retained eight total scores but seven knowledge-transfer scores for the same reason at the category level. Scores with no eligible items were excluded.

In Digital Painting, Preview ($M=75.2\%$, $SD=12.6$) outperformed Video Gen ($M=54.2\%$, $SD=16.2$; $\Delta=21.0$, Welch's $t(13.21)=2.90$, $d\approx1.45$, $p_{\mathrm{adj}}=.0369$). AR scored $M=66.7\%$ ($SD=21.4$). Neither Preview--AR ($d\approx0.48$) nor Video Gen--AR ($d\approx-0.66$) was significant (both $p_{\mathrm{adj}}=.421$). Holm correction covered the three pairwise comparisons within each task. In a sensitivity analysis treating all nine task-specific quiz contrasts as one family, the Preview--Video Gen contrast had adjusted $p=.111$. Exact-permutation checks gave adjusted $p=.0387$ within painting and $.116$ across nine contrasts. Figure~\ref{fig:painting-understanding-execution} pairs painting quiz and execution outcomes, marking significant task-wise contrasts.

Neither Cocktail Making nor Gameplaying showed a significant pairwise contrast under task-wise correction. Their mean accuracies in Preview/Video Gen/AR order were 60.0/62.9/56.9\% and 49.3/62.6/57.5\%, respectively. Shared guidance and task experience also provided sources for quiz answers across conditions. The quiz assessed conceptual and transfer knowledge as well as recall, making procedural understanding distinct from ease of execution. Significant and nonsignificant within-task contrasts alone do not establish that condition effects differ across tasks.

\subsubsection{\textbf{System Usability and Cognitive Workload.}}
\begin{figure*}[htbp]
  \centering
  \includegraphics[width=\linewidth]{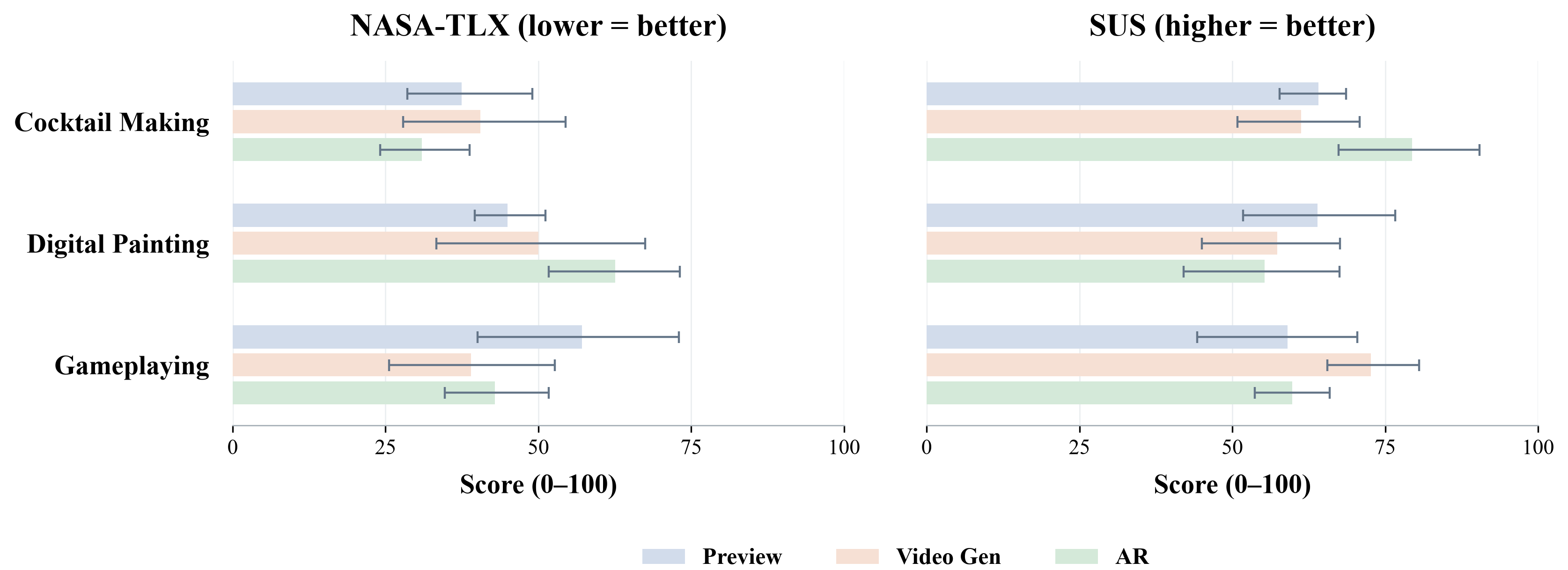}
  \caption{Adapted NASA-TLX (left; lower indicates less workload) and SUS (right; higher indicates greater usability) across three tasks. Preview, Video Gen, and AR correspond to C, B, and A. Bars show means with pointwise 95\% participant-bootstrap CIs (n=8). No within-task contrast was significant after Holm correction.}
  \Description{Two grouped horizontal bar-chart panels show workload and usability for all three tasks. Preview has descriptively lower workload and higher usability than AR in painting, AR has the highest mean usability in cocktail making, and Video Gen has the highest mean usability in gameplay. Error bars display uncertainty around each mean.}
  \label{fig:all-tasks-workload-usability}
\end{figure*}

SUS was collected by condition, so its task-specific summaries associate each rating with the task experienced under that condition. No comparison was significant after Holm correction within each task across two instruments and three condition contrasts (minimum adjusted $p=.157$). Digital Painting showed descriptively lower workload and higher usability with Preview than AR (NASA-TLX: 44.9 versus 62.6; SUS: 63.9 versus 55.3). The pattern differed in the other tasks: AR had the highest mean SUS in Cocktail Making (79.4), whereas Video Gen had the highest in Gameplaying (72.7; Figure~\ref{fig:all-tasks-workload-usability}).

\subsubsection{\textbf{Video Skipping and Representation Preferences.}}

We counted explicit video-skip codes over all annotated steps in each video-supported condition. In Gameplaying, these occurred in 76/104 Video Gen steps (73.1\%) and 77/104 Preview steps (74.0\%). Corresponding rates in Cocktail Making were 38/88 (43.2\%) and 28/88 (31.8\%). Neither video-supported painting condition contained an explicit video-skip code. These are step-annotation counts; the remaining steps combine viewed clips and other uncoded viewing states.

Task-specific rankings compared AR cues, ordinary reference images, and generated action video for next-step guidance (1--3; lower is preferred). Among complete rankings, AR was most preferred in Cocktail Making ($M=1.54$) and Gameplaying ($M=1.33$), and reference images in Digital Painting ($M=1.50$). Video ranks were $M=1.69$, $M=2.58$, and $M=1.64$, respectively. Rankings and skip annotations suggest context-dependent use, distinct from preferences for added future-interface components.

\subsubsection{\textbf{Perceived Value of the Choice Interface.}}

On seven-point scales ($n=24$), participants rated alignment with personal priorities ($M=5.13$, $SD=1.08$), perceived control ($M=4.79$, $SD=1.21$), and interruption ($M=3.13$, $SD=1.68$; higher scores indicate greater interruption). Interviews contextualize these experiences of required choice.

\begin{figure*}[htbp]
  \centering
  \includegraphics[width=\linewidth]{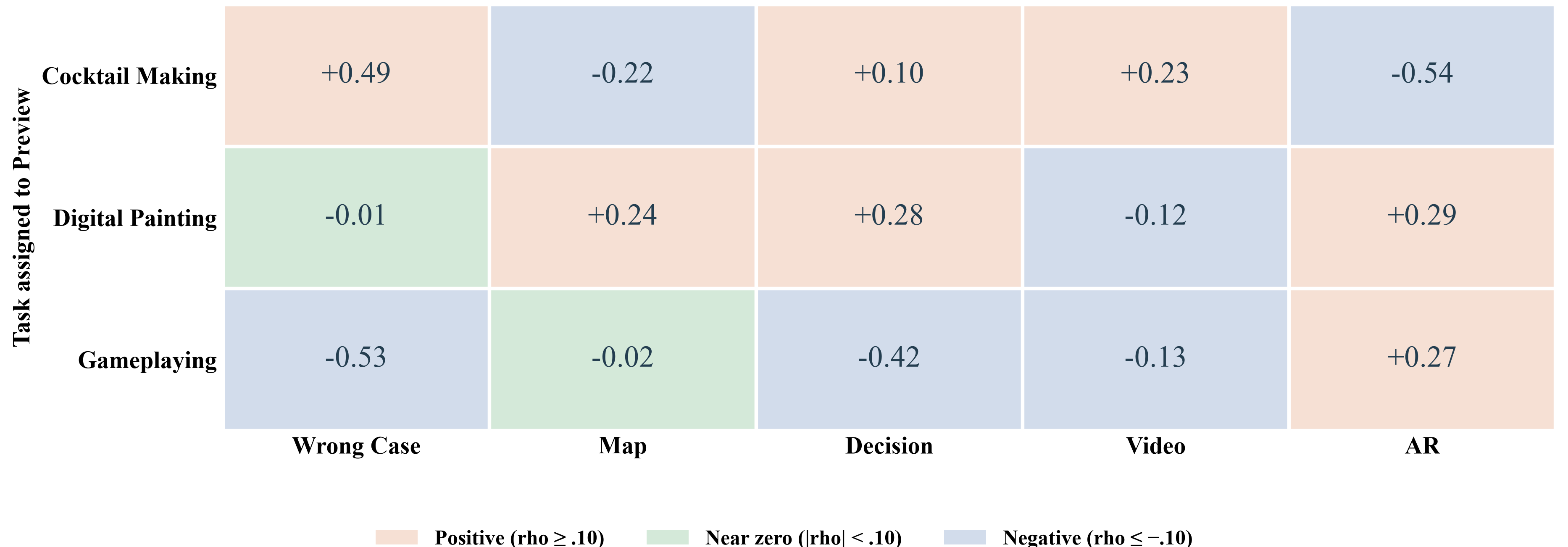}
  \caption{Exploratory Spearman associations ($\rho$) between task assignment in Preview (Condition C) and overall component preference, scored as $6-\mathrm{rank}$. Positive values indicate stronger preference among participants assigned that task in C. Columns identify Wrong Case, Multi-Step Preview, Alternative-Method Preview, Video, and AR. Colors distinguish positive ($\rho\geq .10$), near-zero ($|\rho|<.10$), and negative ($\rho\leq -.10$) coefficients. No association was significant after Holm correction.}
  \Description{A three-by-five heatmap shows task--component association coefficients. Cocktail Making has values of 0.49, minus 0.22, 0.10, 0.23, and minus 0.54; Digital Painting has minus 0.01, 0.24, 0.28, minus 0.12, and 0.29; Gameplaying has minus 0.53, minus 0.02, minus 0.42, minus 0.13, and 0.27.}
  \label{fig:task-component-associations}
\end{figure*}

We additionally examined the association between overall component preference and the task assigned to Preview (Figure~\ref{fig:task-component-associations}). Tasks used binary assignment indicators; preference was $6-\mathrm{rank}$. Positive Spearman coefficients indicate stronger overall preference among those assigned that task in Preview. None of the associations survived Holm correction (minimum adjusted $p=.394$). These associations relate overall preferences to task assignment in Preview, with the other task--condition assignments co-varying.

\subsection{Qualitative Results}\label{sec:qual-results}

The study collected recordings, interviews, think-aloud accounts, and observation notes. One author synthesized interview notes into six exploratory RQ1 themes after inspecting results. The themes paraphrase participants' interpretations and perceived usefulness, retaining contrasting experiences and distinguishing reported from hypothetical use.

\subsubsection{A Visible Outcome Still Needs a Link to the Current Action}\label{sec:qual-action-relevance}

U9 described consulting Deviation\textendash{}Consequence Preview to check an operation when the demonstration video moved too quickly; U13 associated checking it with avoiding a comparable mistake. In contrast, U5 did not recognize its connection to their own problem and wanted specific information about what to change, such as a value or selection. A depicted consequence could support checking without necessarily explaining how to avoid it.

U7 and U14 described inspecting additional information when it matched their situation or uncertainty, while U8 associated inspection with failure or an unexpected result. Preventive information could thus become relevant through an already salient concern. Users could understand the warning's hypothetical status yet see little reason to inspect it. The central issue was linking a possible consequence to an action they could check or change now.

\subsubsection{Richer Guidance Addresses an Unresolved Information Need}\label{sec:qual-information-need}

U8 identified shaking motion and painting trajectories as information video could clarify. U14 turned to video when images and instructions were unclear but considered it unnecessary when AR was sufficiently precise. U22 likewise described proceeding without waiting when an image already conveyed how to act. Demand for detail depended on what remained unresolved at the step.

Familiarity and expected value shaped willingness to wait. U12 distinguished familiar tasks, where text sufficed, from unfamiliar tasks requiring further guidance. U18 considered waiting acceptable because generation seemed relatively fast and the information valuable. These contrasts connect selective inspection to familiarity and expected informational value.

\subsubsection{Knowing Upcoming Steps Does Not Explain Their Connections}\label{sec:qual-map}

U18 associated Multi-Step Preview and the directly displayed target preview with anticipating what would happen next. Other participants sought relationships rather than sequence alone: U14 wanted the logic connecting steps explained, and U23 wanted an account of intermediate painting operations, including layer settings and differences between methods.

\subsubsection{Choice Supports Trade-Offs and Control Under Uncertainty}\label{sec:qual-choice}

Both methods at the decision points pursued the task goal. U22 considered whether an initially convenient action would incur later costs, while U9 preferred a route they could understand and anticipate. Conversely, U7 questioned the value of choosing between routes to the same goal, and U5 described arbitrary selection. Feasible alternatives did not always provide a meaningful basis for comparison.

Even favorable accounts distinguished choice from understanding: U22 felt that choices clarified the goal while remaining unsure what distinguished the routes. U20 valued retaining choice because they did not fully trust the system. Thus, control could coexist with uncertainty about the alternatives or recommendation. The required choice made the basis for comparison especially important.

\subsubsection{Operational Information Determines Whether Generated Content Is Usable}\label{sec:qual-operational-information}

All main-study interviewees reported that hallucinations in generated videos did not impede their understanding of the demonstrated actions. Several nevertheless noted inaccuracies in initial GUI target localization.

U8 described using common sense to interpret an unnatural lid-opening depiction. U18 reported compensating for an offset position using an earlier correct example and AR cues. Requests for clearer brush identification (U18) and slider direction (U20) further highlighted details needed to act. These accounts concern identifying and interpreting the intended operation.

U13 described checking imagery against text, objects, and common sense, whereas U7 trusted the system by default. A plausible depiction could therefore receive different levels of scrutiny. These accounts emphasize identifiable targets, directions, and state changes, with complementary cues helping users interpret imperfect outputs.

\subsubsection{Matching a Visible Result Can Leave Later Possibilities Unclear}\label{sec:qual-rationale}

U5 paid little attention to intermediate painting operations when the visible result looked the same. U9 reported difficulty understanding independent editing in the Digital Painting task, and U20 consulted text and video while still finding consequences unclear. These accounts distinguish matching an appearance from understanding the later operations it permits. Related uncertainty extended beyond painting: U7 questioned why soda water was added later, while U22 discovered an earlier mistake through subsequent steps.

In gameplay, many participants described understanding what each step accomplished as a major difficulty. They reported following bounding boxes and arrows while overlooking accompanying text, videos, and pictorial icons. Their accounts highlight a gap between locating the next action and understanding its purpose.

U21 described a preview becoming meaningful through practical experience only after it had disappeared. This mismatch between presentation and interpretation motivates continued access as users gain experience, alongside separate assessment of execution and understanding.

\subsection{Summary Across Research Questions}

For \textbf{RQ2}, pooled comparisons detected no significant condition differences. Exploratory painting comparisons favored Foresee over AR for correct completion without a recorded error and over Video Gen for quiz accuracy under task-wise correction. For \textbf{RQ1}, participants' accounts highlight unresolved information needs, connections to current and later actions, and opportunities to inspect or revisit guidance. These accounts suggest conditions that may shape preview use, not causal explanations of the quantitative differences.

\section{Design Implications for Making Future Previews Usable}\label{sec:design-implications}

The formative findings and main-study interviews motivate four interaction recommendations and one generation recommendation (Sections~\ref{sec:formative-results} and~\ref{sec:qual-results}). Their central concern is how a preview connects a current action, a possible state, and a later task operation. The examples below translate this concern into design choices, including extensions to Foresee.

\subsection{Interaction Implications}

\subsubsection{Anchor a Visible Consequence to an Action Users Can Check Now}

Participants valued previews without necessarily opening them, and directly visible imagery could still appear unrelated to their situation (Section~\ref{sec:qual-action-relevance}). At an error-sensitive step, pair the consequence image with the action and task requirement that make it relevant. For example, ``If you draw the sun on the background layer, moving it later may also affect the background'' identifies both the current choice and its later cost. Keep this concise relation visible beside the instruction, with further inspection optional. Label the depicted deviation as hypothetical so users can distinguish a possible consequence from feedback about an error they have already made.

\subsubsection{Show Later Capabilities, Not Only Future Appearance}

Unresolved concepts such as independent editing show why a plausible outcome image may leave an action's value unclear (Sections~\ref{sec:qual-map} and~\ref{sec:qual-rationale}). Select a later operation that exposes the difference: two drawings may look identical until the user tries to move one element independently. A comparison could hold the initial appearance constant and depict what happens when that operation is attempted. Likewise, Multi-Step Preview could connect layer creation to later editing through an annotated dependency, rather than presenting the steps only in sequence. Choose an image, animation, or explanation according to what makes this relationship legible.

\subsubsection{Separate Immediate Action Detail from Revisitable Explanations}

Participants' selective waiting distinguishes an immediate question about execution from a question about consequences (Section~\ref{sec:qual-information-need}). Offer concrete motion detail when users need to see a drag, stroke, or shaking action; label access by the question answered, such as ``Show the shaking motion.'' Keep the instruction and available imagery accessible while video loads. For gameplay, where participants followed spatial markers while overlooking explanations, brief optional speech could connect the highlighted action to its purpose at a suitable pause. Replay and mute controls would let users regulate this proposed form of support.

Multi-Step Preview already provides access to earlier content. U21's account of understanding a preview only after acting motivates making that return path easier to discover (Section~\ref{sec:qual-rationale}). Identify the originating step and captured state when content is revisited or arrives late, so users can relate it to their experience without confusing it with guidance for the active step.

\subsubsection{Compare Methods by Their Downstream Trade-Offs}

Required choice sometimes supported control while leaving the difference between methods unclear (Section~\ref{sec:qual-choice}). Organize each comparison around the later operation or cost that distinguishes the routes. A painting panel could ask whether an element will need independent movement or recoloring, then explain how each method supports that priority. When the goal favors one route, provide a reasoned recommendation with access to the alternative. When both remain suitable, let users decide which consequence matters to them. Optional comparison and goal elicitation would extend Foresee's fixed, required branch choices.

\subsection{Preserve Action--Consequence Consistency Across Generated Media}

Participants compensated for some visual artifacts using text, spatial cues, and common sense, while requesting clearer targets and directions (Section~\ref{sec:qual-operational-information}). Generation checks should therefore address the instructional relationship across representations: the named target, depicted action, resulting state, and stated consequence should agree. For a GUI operation, proposed checks could verify that the named control exists and that the indicated point lies within its region, then examine whether the depicted transition matches the action. Failed checks could trigger review or fallback to reviewed guidance. Structured rendering supplies control over rule-defined content; image and video generation supply appearance and motion, with checks at their boundaries preserving the intended operation.

\section{Discussion}\label{sec:discussion}

Foresee brings action demonstrations and prospective explanations into the same procedural interface. The findings highlight three distinctions: making information available versus making it relevant, depicting appearance versus explaining capability, and supplying detail versus creating an opportunity to interpret it.

\subsection{From Inviting Reflection to Making a Consequence Interpretable}

\emph{From Following to Understanding} demonstrates the value of reflective prompts and discusses balancing reflection with task flow~\verifiedcite{zhang2025reflective}. Foresee examines a complementary representational question: how do users interpret a supplied depiction of an action's consequences? The formative study identified barriers to accessing valued information, motivating direct presentation. The main study exposed a further challenge: visible consequences could still lack a clear relation to the current action. Access and interpretation are therefore distinct design concerns.

This progression shifts the design focus from presenting more future information to making its procedural relevance inspectable. A depicted state needs a connection to the current action and the task requirement that gives it meaning. The same principle applies to hypothetical errors and feasible alternatives: their significance lies in what they imply for the user's goal, not merely in being visible or selectable.

\subsection{A Future Appearance Can Conceal a Difference in Capability}

A generated image can make an expected appearance concrete while leaving its procedural significance implicit. Participants' accounts distinguish recognizing a state from understanding which later actions it permits (Section~\ref{sec:qual-rationale}). This motivates treating a procedural future as an action--state--action relationship: the later operation reveals why the state produced by the current action matters.

The quantitative comparisons concern guidance packages, not isolated explanatory mechanisms. They do not establish that understanding later capabilities caused better performance. Rather, the interviews motivate a representational direction for further testing: depict what a state enables or constrains, alongside how to produce it. This extends outcome previewing from showing an expected appearance to explaining a procedural dependency.

\subsection{Waiting and Attention Shape the Value of a Preview}

Selective viewing suggests that detail has value relative to an unresolved question, rather than through a fixed hierarchy of media. Images and spatial cues may suffice for orientation, while video supplies motion that remains difficult to infer. Willingness to wait reflects the expected value of that additional information.

Non-use is equally important to interpret cautiously. Gameplay accounts combined following spatial cues with difficulty understanding step purposes. Skipping a preview may therefore reflect either sufficient action guidance or limited attention despite an explanatory need. Ease of progression should not be treated as evidence of understanding, nor low inspection as evidence that explanations lack value.

Reflective and proactive AR research considers how assistance fits ongoing activity~\verifiedcite{zhang2025reflective,satori25}. Foresee adds a distinction between when a representation becomes available and when users can interpret it. Generation time is only one part of this interval; attention and practical experience also shape opportunities for understanding. Future evaluation should therefore consider both immediate action support and later explanatory use, rather than treating first presentation as the only opportunity for a preview to be useful.

\section{Limitations and Future Work}\label{sec:limitations}

\subsection{Study Scope and Task-Specific Findings}

The framework and six-participant formative study draw on selected configurations and evolving prototypes. The main study involved 24 participants, with eight per task--condition cell and each participant experiencing every condition on a different task. The task-specific comparisons are exploratory, and the painting quiz finding depends on the multiplicity policy. Larger studies could examine how task familiarity, visual demands, and opportunities to pause shape the value of previews. The qualitative synthesis was led by one researcher after examining the results; further studies across users and procedures could refine these interpretations.

\subsection{Separating Content, Presentation, and Interaction}

Foresee combines outcome images, explanatory text, cross-step access, and required method choices. Conditions varied both explanatory content and presentation, so the findings concern these guidance packages. Matched-content comparisons could examine the contribution of each representation and identify a smaller sufficient feature set. Comparing required and optional method choices could also clarify when choosing promotes reflection and when it interrupts execution. For alternatives with the same final goal, useful outcomes include understanding the trade-off and choosing a method aligned with an independently elicited preference.

\subsection{Attention, Understanding, and Learning Over Time}

The quizzes assessed post-task factual, conceptual, and transfer knowledge using participant-specific item sets that excluded self-reported prior-known items. Because questionnaires and quizzes followed all trials, responses may reflect recall and intervening experiences. Pre-task, immediate, and delayed assessments could trace learning and retention, while modified tasks could test transfer. Linking preview-viewing logs with actions and explanations would clarify when participants consult guidance and how their interpretations change through execution.

Gameplay participants' focus on boxes and arrows, alongside difficulty understanding step purposes, motivates support that fits ongoing visual activity. Future work could compare visual explanations with brief optional spoken explanations at suitable pauses, assessing comprehension, interruption, and user control. Persistent access could likewise support revisiting an explanation after practical experience makes it meaningful.

\subsection{GUI Localization Accuracy and Human Assistance}

GUI target localization and screen registration constrain the spatial accuracy of generated guidance. Participants reported initial localization errors, and incorrect target positions can propagate into misplaced action demonstrations. ScreenSpot-Pro similarly identifies difficulties with small targets and icons in high-resolution professional interfaces~\cite{li2025screenspotpro}. In our study, backend GUI-target assistance was available across conditions in painting and gameplay, making human assistance part of the evaluated workflow. Future evaluation should measure target accuracy before correction, record the frequency and duration of assistance, and trace target positions into generated outputs. This would help distinguish errors in perception and alignment from artifacts introduced during media generation.

\subsection{Generation Quality and Runtime Performance}

Generated-video quality was examined through observations and interviews, with interviewees reporting that hallucinations did not impede their understanding of the demonstrated actions. A complementary content assessment could examine action correctness, target identity, and action--outcome consistency across generated outputs. Comparisons with matched prerecorded media and alternative generation pipelines could assess the value of scene adaptation alongside authoring effort and the review of procedural plans.

The runtime analysis summarizes historical service requests, including development tests, and measures server processing time. Participant- and session-linked logging could connect individual user requests to intermediate jobs, retries, assistance, and the time at which guidance becomes visible. Such measurements would capture how end-frame preparation and generation scheduling affect the wait users experience, alongside service reliability.

\section{Conclusion}

We presented Foresee, a generative future interface that pairs action demonstrations with consequence previews, method comparisons, and cross-step access. A literature-informed framework and six-participant formative study shaped its design, implemented through a hybrid pipeline combining structured rendering, image editing, and video generation. In a 24-participant study across three task domains, exploratory painting comparisons showed higher correct completion without a recorded error than AR and higher post-task quiz accuracy than generated-video guidance under task-wise correction. Interviews highlighted a distinction between following an action and understanding what it enables: participants could overlook available explanations, match visible outcomes while missing procedural relationships, or wait for concrete demonstrations when symbolic cues were unclear. These findings motivate previews that connect consequences to current actions, explain later capabilities, and provide detail at moments users can attend to it. Foresee contributes an interface and design guidance for making procedural futures understandable within ongoing task execution.

\bibliographystyle{ACM-Reference-Format}
\bibliography{software}

\end{document}